\documentclass[10pt,journal,compsoc]{IEEEtran}

\IEEEoverridecommandlockouts
\usepackage{graphicx}

\ifCLASSOPTIONcompsoc
    \usepackage[caption=false, font=normalsize, labelfont=sf, textfont=sf]{subfig}
\else
\usepackage[caption=false, font=footnotesize]{subfig}
\fi
\usepackage{cite}
\usepackage{url}
\usepackage{mdwmath}  
\usepackage{soul}
\usepackage[table]{xcolor}
\usepackage{amsfonts,amsmath,amssymb,amsxtra,amsbsy,floatflt}
\usepackage{indentfirst} 
\usepackage{tabularx}
\usepackage{booktabs}
\usepackage{multirow}
\usepackage{nth}
\usepackage{comment}
\usepackage{epstopdf}
\usepackage{hhline}
\usepackage{siunitx}
\usepackage{booktabs}
\usepackage[bottom]{footmisc}
\usepackage[acronym]{glossaries}
\usepackage{upgreek}
\usepackage{algorithm}
\usepackage{algorithmicx,algpseudocode}
\usepackage{bm}
\usepackage{upgreek}
\usepackage[normalem]{ulem}
\usepackage{arydshln}

\newtheorem{problem}{Problem}
\newcommand{\vv}[1]{\boldsymbol{\mathbf{#1}}}

\newcommand{\tr}{\mathrm{tr}}

\newcommand{\tran}{\mathrm{T}}
\newcommand{\herm}{\mathrm{H}}
\newcommand{\Exp}{\mathbb{E}}

\newcommand{\ssub}[1]{{\scriptscriptstyle { #1}}}
\newcommand{\change}[1]{{\color{black} {#1}}}

\newcommand{\name}{{ARES}}
\newcommand{\oldname}{{MARISA}}

\newacronym{3d}{3D}{three dimensional}
\newacronym{aoa}{AoA}{angle of arrival}
\newacronym{aod}{AoD}{angle of departure}
\newacronym{ap}{AP}{access point}
\newacronym{b5g}{B5G}{Beyond-5G}
\newacronym[plural=BSs, firstplural=base stations (BSs)]{bs}{BS}{base station}
\newacronym{csi}{CSI}{channel state information}
\newacronym{dc}{DC}{direct current}
\newacronym{dl}{DL}{downlink}
\newacronym{doa}{DoA}{direction-of-arrival}
\newacronym{dsp}{DSP}{digital signal processing}
\newacronym{emf}{EMF}{electromagnetic field}
\newacronym{em}{EM}{electromagnetic}
\newacronym{fp}{FP}{fractional program}
\newacronym[plural=HRISs, firstplural=Hybrid Reconfigurable Intelligent Surfaces (HRISs)]{hris}{HRIS}{hybrid reconfigurable intelligent surface}
\newacronym{in-out}{I/O}{input-output}
\newacronym{ios}{IoS}{Internet-of-Surfaces}
\newacronym{iot}{IoT}{Internet-of-Things}
\newacronym[plural=KPIs, firstplural=key performance indicators (KPIs)]{kpi}{KPI}{key performance indicator}
\newacronym{lf}{LF}{low frequency}
\newacronym{los}{LoS}{line-of-sight}
\newacronym{mc}{MC}{Markov chain}
\newacronym{mimo}{MIMO}{multiple-input multiple-output}
\newacronym{mmimo}{M-MIMO}{massive-MIMO}
\newacronym{miso}{MISO}{multiple-input single-output}
\newacronym{ml}{ML}{machine learning}
\newacronym{mmse}{MMSE}{minimum mean squared error}
\newacronym{mrt}{MRT}{maximum-ratio transmission}
\newacronym{mse}{MSE}{mean squared error}
\newacronym{nlos}{NLoS}{non-line-of-sight}
\newacronym{pdf}{pdf}{probability distribution function}
\newacronym{pin}{PIN}{positive-intrinsic-negative}
\newacronym{pla}{PLA}{planar linear array}
\newacronym{pap}{P\&P}{plug-and-play}
\newacronym{ppp}{PPP}{Poisson point process}
\newacronym[plural=RISs, firstplural=Reconfigurable Intelligent Surfaces (RISs)]{ris}{RIS}{Reconfigurable Intelligent Surface}
\newacronym{rf}{RF}{radio frequency}
\newacronym{rmse}{RMSE}{root-mean-square error}
\newacronym{rss}{RSS}{received signal strength}
\newacronym{rv}{RV}{random variable}
\newacronym{sdp}{SDP}{semidefinite programming}
\newacronym{sdr}{SDR}{semidefinite relaxation}
\newacronym{sinr}{SINR}{signal-to-interference-plus-noise ratio}
\newacronym{smse}{SMSE}{sum mean squared error}
\newacronym{snr}{SNR}{signal-to-noise ratio}
\newacronym{soa}{SoA}{state-of-the-art}
\newacronym{soc}{SoC}{state-of-charge}
\newacronym{sre}{SRE}{smart radio environment}
\newacronym{tdd}{TDD}{time division duplexing}
\newacronym{toa}{ToA}{time-of-arrival}
\newacronym[plural=UEs, firstplural=user equipments (UEs)]{ue}{UE}{user equipment}
\newacronym{ul}{UL}{uplink}
\newacronym{ula}{ULA}{uniform linear array}
\newacronym{eh}{EH}{energy harvesting}
\newacronym{ran}{RAN}{radio access network}
\newacronym{cdf}{CDF}{cumulative distribution function}
\newacronym{loc}{LoC}{loss of charge}

\title{ARES: Autonomous RIS solution with Energy harvesting and Self-configuration towards 6G}
\author{Antonio~Albanese,~\IEEEmembership{Member,~IEEE,}
Francesco~Devoti,~\IEEEmembership{Member,~IEEE,}
Vincenzo~Sciancalepore,~\IEEEmembership{Senior Member,~IEEE,}
Marco~Di~Renzo,~\IEEEmembership{Fellow,~IEEE,} Albert Banchs, \IEEEmembership{Senior Member,~IEEE,}
Xavier~Costa-P\'erez,~\IEEEmembership{Senior Member,~IEEE}
\thanks{\textit{A. Albanese is with NEC Laboratories Europe, 69115 Heidelberg, Germany, and Flyhound Co., 10019 New York, USA, and University Carlos III of Madrid, 28911 Legan\'es, Spain. 
F. Devoti and V. Sciancalepore are with NEC Laboratories Europe, 69115 Heidelberg, Germany. 
M. Di Renzo is with Universit\'e Paris-Saclay, CNRS, CentraleSup\'elec, 91190 Gif-sur-Yvette, France. Albert Banchs is with University Carlos III of Madrid, and IMDEA Networks Institute, 28911 Legan\'es, Spain.
X. Costa-P\'erez is with i2cat, and ICREA, and NEC Laboratories Europe, 08034 Barcelona, Spain. 
This work was supported by EU H2020 RISE-6G project under grant agreement no. 101017011. Email of the corresponding author: francesco.devoti@neclab.eu.}}%
}

\begin{document}

\maketitle

\begin{abstract}
\change{\Glspl{ris} are expected to play a crucial role in reaching the \glspl{kpi} for future 6G networks. Their competitive edge over conventional technologies lies in their ability to control
the wireless environment propagation properties at will,
thus revolutionizing the traditional communication paradigm that perceives the communication channel as an uncontrollable black box. As \glspl{ris} transition from research to market, practical deployment issues arise. Major roadblocks for commercially viable \glspl{ris} are $i$) the need for a fast and complex control channel to
adapt to the ever-changing wireless channel conditions, and $ii$) 
an extensive grid to supply power to each deployed \gls{ris}.
In this paper, we question the established \gls{ris} practices and propose a novel \gls{ris} design combining self-configuration and energy self-sufficiency capabilities. We analyze the feasibility of devising \emph{fully-autonomous \glspl{ris}} that can be easily and seamlessly installed throughout the environment, following the new \gls{ios} paradigm, requiring modifications neither to the deployed mobile network nor to the power distribution system. In particular, we introduce \name{}, an \textit{Autonomous \gls{ris} with Energy harvesting and Self-configuration solution.} 
\name{} achieves outstanding communication performance while demonstrating the feasibility of \gls{eh} for \glspl{ris} power supply in future deployments.}
\end{abstract}


\glsresetall
\vspace{-0.2cm}
\section{Introduction}
\label{sec:intro}

\change{The ever-increasing need for network performance improvement has led to exploring new ways to push communication transmission efficiency that go beyond the classic communication paradigm. A revolutionary technology capable of providing control over the propagation environment has been recently introduced: metasurfaces with their reflectarray-based variant, namely \gls{ris}, raise the possibility of altering the way waves propagate in the environment in an intelligent, controllable and flexible fashion, opening up the possibility of optimizing it for unprecedented communication performance~\cite{TAOM20}.
While this game-changing technology introduces a bulk of new business opportunities and advanced use-cases for the next generation of wireless networks (B5G or 6G), it involves new challenges to be addressed~\cite{RZDAC_jsac20}.

The \gls{ris} paradigm transforms the propagation environment from an adversary to an optimizable communication ally actively contributing to improving performance, with the sole use of surfaces equipped with low-cost, and low-complexity electronics~\cite {RIScommag_2021}. However, \glspl{ris} require an ad-hoc control channel to adapt to the network dynamics. Furthermore, being quasi-passive devices, channel estimation requires complex procedures performed on the entire transmitter-\gls{ris}-receiver path, which can hamper their agile deployment~\cite{di2019smart}.
To counter this drawback, the concept of \gls{hris} has been proposed in~\cite{alexandropoulos2021hybrid}, bringing built-in sensing capabilities to quasi-passive \glspl{ris}. This kick-started the development of self-configuring \glspl{ris} as the epitome of a \gls{pap} \glspl{ris} integration solution into existing network deployments that eliminates the need for the control channel and operator management, paving the way to the massive and flexible deployment of \glspl{ris} known as \gls{ios}~\cite{albanese2022marisa}.

As operational \glspl{ris} draw limited amount of energy~\cite{rossanese2022}, they may run on compact batteries, thereby requiring no continuous external power supply but calling for periodic maintenance, which might limit their operational availability. Alternatively, \gls{eh} techniques may be implemented at the \glspl{ris} to provide them with potentially unlimited energy availability, avoiding the need for servicing for long time periods~\cite{zeadally2020design} and leading to a fully autonomous solution in terms of configuration, energy, and maintenance.

Pursuant to the \gls{ios} framework, we select \glspl{hris} as reference hardware due to their signal reflection, absorption, and sensing capabilities. In~\cite{albanese2022marisa}, we previously analyzed one of the fundamental constraints of a conventional \glspl{ris} deployment, i.e. the need for a reliable control channel, and introduced \oldname{} as a novel system design that achieves fully-autonomous and self-configuring \glspl{ris} deployments. In this paper, we extend~\cite{albanese2022marisa}, going one step beyond and targeting the need for a power supply to devise a \textit{fully-autonomous}, \textit{self-configuring}, and \textit{energy self-sufficient} metasurface solution, namely Autonomous RIS with Energy harvesting and Self-configuration (\name{}).
\name{} takes advantage of the key assets brought in by \oldname{}, namely the new channel estimation model lato-sensu at the \gls{hris} and the autonomous \gls{hris} configuration methodology based only on the \gls{csi} available at the \gls{hris}, and establishes the following main contributions: $i$) the design of a fully integrated \gls{rf} \gls{eh} technology to empower off-the-grid operations, $ii$) an accurate model of the battery charging and discharging processes based on an irreducible \gls{mc}, $iii$) a battery dimensioning strategy that guarantees the availability of the \gls{hris} for the desired time period.
\name{} is shown to provide near-optimal performance when compared to the full-\gls{csi}-aware and power-grid-enabled approaches while facilitating energy self-sufficient \gls{ios} deployments.}

\change{The remainder of this paper is as follows. Section~\ref{sec:hris} provides preliminary information about our \gls{hris}-based design pointing out the major roadblocks to a fully-autonomous \gls{ios} deployment. Section~\ref{sec:framework_overview} provides an overall view of the building blocks of our solution. Section~\ref{sec:model_design} introduces the system model and the optimization problem for an \gls{hris} without \gls{rf} chains and a dedicated control channel. Besides, it presents a novel analysis for energy management at the \gls{hris} leveraging on a bespoke Markov-chain model to represent the battery charge and discharge processes. Section~\ref{sec:absorption_model} unlocks the \gls{hris} self-configuration and energy self-sufficiency by tackling the aforementioned problem via codebook-based optimization, which is complemented by the assessment of the \gls{hris} hardware-related power consumption. Section~\ref{sec:results} shows \name{} performance against \gls{soa} and heuristic solutions whereas Section~\ref{sec:related} includes a survey of the related literature. Finally, Section~\ref{sec:conclusion} concludes the paper.}

\emph{Notation}. We denote matrices and vectors in bold text while each of their element is indicated in roman with a subscript. $(\cdot)^{\tran}$ and $(\cdot)^{\herm}$ stand for vector or matrix transposition and Hermitian transposition, respectively. The L$2$-norm of a vector is denoted by $\| \cdot \|$ while $\tr(\cdot)$ indicates the trace of a matrix. Also, $\langle\cdot, \cdot \rangle$ denotes the inner product between vectors, and $\circ$ denotes the \emph{Hadamard} product between two matrices.

\vspace{-0.19cm}
\section{Related Work}
\label{sec:related}

\change{In the last few years, \glspl{ris} have drawn considerable interest from the scientific community due to their ability to turn uncontrollable propagation channels into controlled variables that can be optimized~\cite{LZWXA21_LCOMM,LFY21_Access}. A preliminary analysis of the achievable performance of an \gls{ris} is given in~\cite{WZ2019_TWC}. In particular, the authors formulate a joint optimization problem for optimizing the active beamforming (at the multi-antenna \gls{bs}) and the passive beamforming configuration (at the \gls{ris}), and they demonstrate that \gls{ris}-based \gls{mimo} systems can achieve rate performance similar to legacy massive \gls{mimo} systems with fewer active antenna elements. The ideal case study with continuous phase shifts at the \gls{ris} is generalized to the case with discrete phase shifts in~\cite{WZ2020_TCOM}. The authors prove the squared power gain with the number of reflecting elements even in the presence of phase quantization, but a power loss that depends on the number of phase-shift levels is observed~\cite{yub21_TCOM}. In~\cite{Mursia2021}, the authors propose a practical algorithm to maximize the system sum mean squared error while jointly optimizing the transmit beamforming at \gls{mimo} \glspl{bs} and the \gls{ris} configuration. 
Other papers have recently considered the possibility of optimizing the \glspl{ris} based on statistical \gls{csi} in order to relax the associated feedback overhead. A two-timescale transmission protocol is considered in~\cite{zhao2020intelligent} to maximize the achievable average sum-rate for an \gls{ris}-aided multiuser system under a general correlated Rician channel model, whereas~\cite{ADD21_TCOM} and \cite{zprw21_LWC} maximize the network sum-rate by means of the statistical characterization of the locations of the users, which does not require frequent updates of the \gls{ris} reconfiguration. These solutions, however, rely on the presence of a control channel.
A detailed analysis of an \gls{ris}-assisted multi-stream \gls{mimo} system is described in~\cite{PTDF2021_TWC}, where the authors formulate a joint optimization problem of the covariance matrix of the transmitted signal and the \gls{ris} phase shifts. An effective solution is obtained, which offers similar performance to \gls{soa} schemes but with limited computational complexity. The approach is generalized in \cite{PTDF2021_LWC} to discrete-valued constellation symbols. A comprehensive tutorial on \glspl{ris} focused on optimization is available in~\cite{WZZYZ2021_TCOM}.

As for the \glspl{ris} design, several prototypical architectures are available in the literature. The majority of works consist of a \gls{pin} diode-based design, achieving different phase shift quantization levels. Early works like~\cite{li2020} and \cite{Trichopoulos2022} propose \gls{ris} hardware designs with 1-bit phase shift resolution, operating at 28 GHz and 5.8 GHz, respectively. Similar approaches offer more quantization bits at the expense of higher circuital complexity, such as~\cite{dai2020reconfigurable} producing 4 possible phase shifts with 5 \gls{pin} diodes per element and~\cite{Hu2020} allowing for 8 phase shifts by means of 3 \gls{pin} diodes. A multi-frequency design is disclosed in~\cite{maresca2022}, paving the way for the development of \glspl{ris} able to dynamically operate across different carrier frequencies by using frequency-tunable antennas as elements. 
In addition to \gls{pin} diodes, varactor diodes may be employed to control the phase response of each \gls{ris} element, enabling continuous tuning of the applied phase shifts~\cite{fara2021}. As a reference, \cite{Fara2022} proposes to use 4 varactors per element in order to ensure a symmetric radiation pattern in the 5.15-5.75 GHz frequency range. 
Alternative designs make use of \gls{rf}-switches~\cite{tan2018}, thanks to the lower associated production cost. For instance, the authors of~\cite{Arun2020} design and build a two-dimensional surface consisting of 80 rectangular boards, each containing $40$ antennas on custom-printed circuit boards. Specifically, the \gls{rf}-switches determine whether the signal propagates through the surface or is reflected. On these lines, \cite{Dunna2020} proposes a 5 GHz \gls{ris} design with 2-bit phase resolution by leveraging on transmission lines to generate different phase shifts at each \gls{ris} element whereas the \gls{rf}-switches are used to route the signals on the optimal line (according to the \gls{ris} configuration). Besides, the work in~\cite{rossanese2022} builds upon the transmission-line method to design a 3-bit resolution \gls{ris} and introduces the possibility of setting each of the \gls{ris} elements (separately) to fully dissipate the impinging \gls{rf} power, i.e., effectively turning off some of the \gls{ris} elements, and virtually changing the overall shape of the array.

None of the above-mentioned works deal with self-configuring \gls{ris}-empowered networks without relying on a control channel. Moreover, although~\cite{rossanese2022} hints to an absorption mode for \gls{ris} elements, such energy is only dissipated over a matched load and is not used to power the device and enable its energy self-sufficiency, thereby underlining the contributions and novelty of the present paper.}

\vspace{-0.2cm}
\section{Plug \& Play HRIS: Key Characteristics}
\label{sec:hris}

In this section, we introduce the key concept of \gls{hris} and provide a brief overview of the proposed bespoke hardware design. Then, we analyze the challenges that need to be overcome \change{to enable self-configuration and self-sufficiency capabilities for the \gls{hris}.}

\begin{figure}[t]
        \center
        \includegraphics[width=1\linewidth]{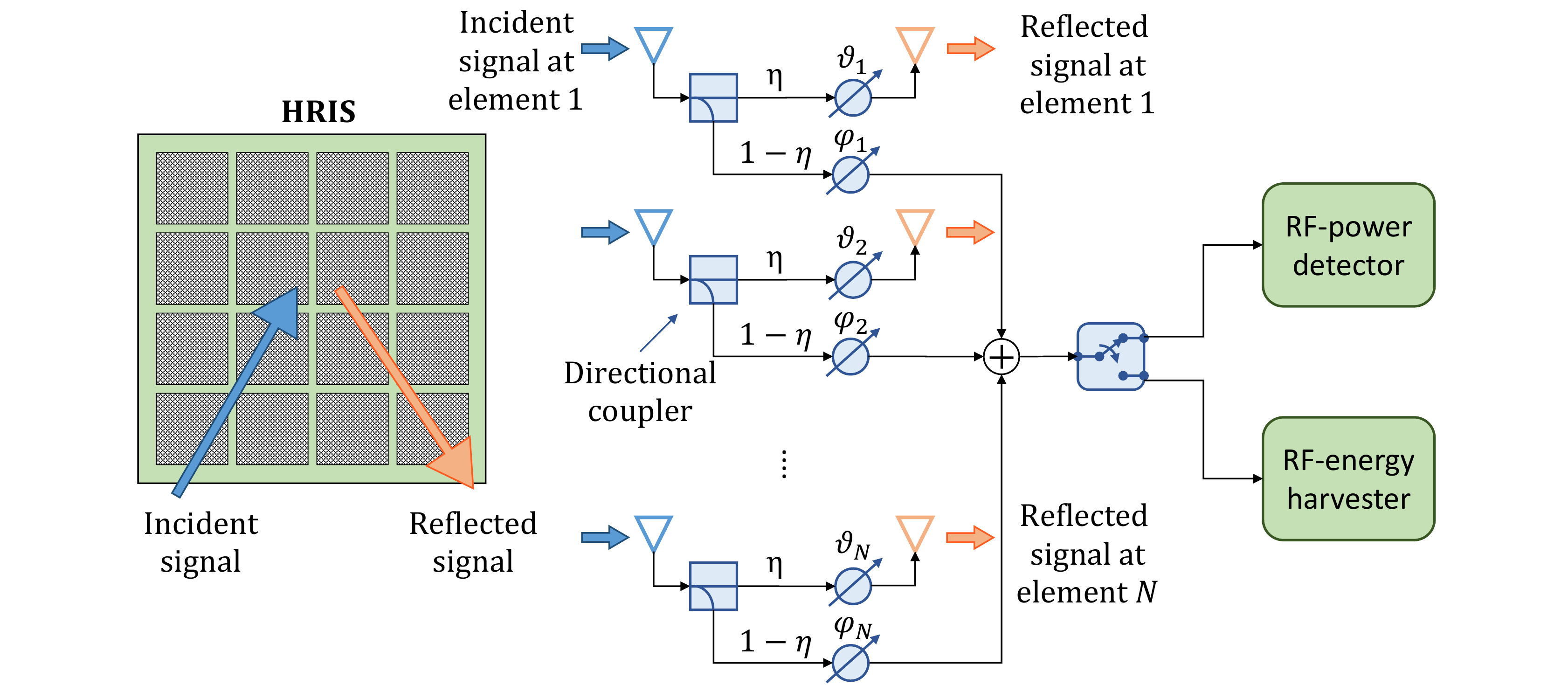}
        \caption{Reference diagram of a hybrid reconfigurable intelligent surface.}
        \label{fig:hris}
\end{figure}

\vspace{-0.3cm}
\subsection{Hardware design}
\label{sec:hw_design}

\textbf{Hybrid metasurfaces.} We consider an \gls{hris}~\cite{alexandropoulos2021hybrid} comprising an array of hybrid meta-atoms, which are able to simultaneously reflect and absorb (i.e., sense the power of) incident signals. In the considered architecture, each metasurface element is coupled with a sampling waveguide that propagates the absorbed (i.e., sensed) power of the incident \gls{em} waves towards some downstream \gls{rf} hardware for enabling \gls{dsp}. 
\change{To reduce the complexity and cost of the required hardware with respect to conventional \gls{hris} and enable \gls{eh}, we propose \gls{pap} \glspl{hris} that are not equipped with fully-fledged \gls{rf} chains but only with an \gls{rf} power detector and an \gls{rf} energy harvester. }
\change{As shown in Fig.~\ref{fig:hris}, the signals absorbed by each metasurface element are summed together by means of \gls{rf} combiners, which may be easily implemented as lumped components throughout the metasurface \gls{rf} circuit~\cite{lamminen200860}. The resulting signal is fed into an \gls{rf} switch, which routes it to either the above-mentioned power detector or the energy harvester. The former converts the \gls{rf} power into a measurable \gls{dc} or a \gls{lf} signal, and is made by, e.g., a thermistor or a diode detector, \cite{li2020novel,yasir2019integration}. The latter extracts energy from the \gls{em} fields in the form of \gls{dc} voltage by means of a rectifier or a voltage multiplier that boosts the output DC by stacking multiple rectifiers, such as in the Cockcroft–Walton or Dickson configurations~\cite{Tran2017}. By leveraging on such components, we respectively enable \glspl{hris} self-configuration and energy self-sufficiency\footnote{\change{In some implementation, the \gls{rf} power detector and the \gls{rf} energy harvester might be fully integrated~\cite{partal2019design}. However, the electronic circuit design of such device is out of the scope of this paper.}}. 

\textbf{Phase shifters banks design.} In the considered hardware architecture, the reflected and absorbed signals are subject to a phase shift applied by the metasurface elements. In particular, each signal is fed to its corresponding phase shifter bank, which is optimized independently of the other, allowing us to simultaneously control the signal reflection and power absorption properties of the \gls{hris}. We will discuss the effects of dependent phase shifts optimization, i.e., the condition in which the same phase shifter bank operates on both the reflected and absorbed signals, in Section~\ref{sec:algorithm}. Here, we underline that independent phase shifter banks can be obtained at once by introducing one additional phase shifter at each meta-atom even starting from a dependent phase shifter banks design, though at the expense of higher circuit complexity.} 

\vspace{-0.3cm}
\subsection{The Road Towards the IoS: Self-Configuring and Self-Sufficient RISs}
\label{sec:hris_key_charac}

\change{\textbf{Managed \glspl{ris} deployment.}} Conventional \gls{soa} \glspl{ris} deployments rely on a control channel between the \glspl{ris} and a centralized controller\footnote{\glspl{ris} are conventionally controlled by a centralized entity (i.e., an orchestration layer) or by the \gls{bs} itself.
}, which serves a twofold purpose: $i$) sharing the \gls{csi} estimated at the \gls{bs} and the \glspl{ris}, $ii$) enabling the joint optimization of the \gls{bs} precoding matrix and the phase shifts at the \gls{ris} elements, in order to avoid losses due to the out-of-phase reception of \gls{ul} signals at the \gls{bs} or \gls{dl} signals at the \acrfullpl{ue}. Indeed, if both the direct and reflected (through one or multiple \glspl{ris}) propagation links between the \gls{bs} and a \gls{ue} are available, the transmission delays experienced by the transmit signals over the two paths may be substantially different, thereby requiring the \gls{ris} to be configured for compensating them. Such configuration is feasible only if the centralized control entity has access to the \gls{csi} of the direct and reflected propagation channels, as well as having full control on the \gls{ris} configuration. \change{Furthermore, although \glspl{ris} operation requires little power (mainly for the \gls{ris} controller and the phase shifting circuitry), currently envisioned \glspl{ris} deployments run on an extensive electrical system, which powers each individual \gls{ris} to enable their wave manipulation functionality~\cite{wu2021intelligent}. Therefore, managed \glspl{ris} deployments $i$) raise doubts as to their cost-effectiveness when compared to \glspl{bs} deployments without \glspl{ris}, thus discouraging their full-scale adoption, $ii$) bring along stringent physical constraints (i.e., need for control channel and electrical power) to network operators willing to be early adopters of such technology~\cite{albanese2021rennes}}.

\change{\textbf{Autonomous \glspl{hris} deployment.} Avoiding the need for an external management-and-control entity and an extensive power grid has major positive implications for the design and deployment of \gls{ris}-aided wireless networks. In the \gls{ios} landscape it is envisioned that novel \glspl{ris} devices
will be completely autonomous in terms of configuration and power supply, requiring neither a dedicated control channel nor an electrical power source, thereby maximizing the agility and flexibility of their deployment while keeping the installation, configuration and maintenance costs minimal.
While typical implementations of \glspl{ris} cannot operate without an external control channel,
the self-configuring \gls{hris} we recently proposed in~\cite{albanese2022marisa} does not rely on the existence of a remote control channel. Instead, it is built upon the optimization and configuration of the \gls{hris} uniquely based on local estimates of the \gls{csi} at the \gls{hris} itself. In this work, we extend our previous work in~\cite{albanese2022marisa} and bring the self-configuring \gls{hris} to the next level, allowing the \gls{hris} to \textit{detach from a continuous power supply} by integrating the capability of harvesting the absorbed \gls{em} energy, which is stored in a properly-sized rechargeable battery to support off-the-grid operations, as depicted in Fig.~\ref{fig:hris_block_diagram}. Hence, we can achieve autonomy both from the control and the energy standpoints.

\begin{figure}[t]
        \center
        \includegraphics[width=1\linewidth]{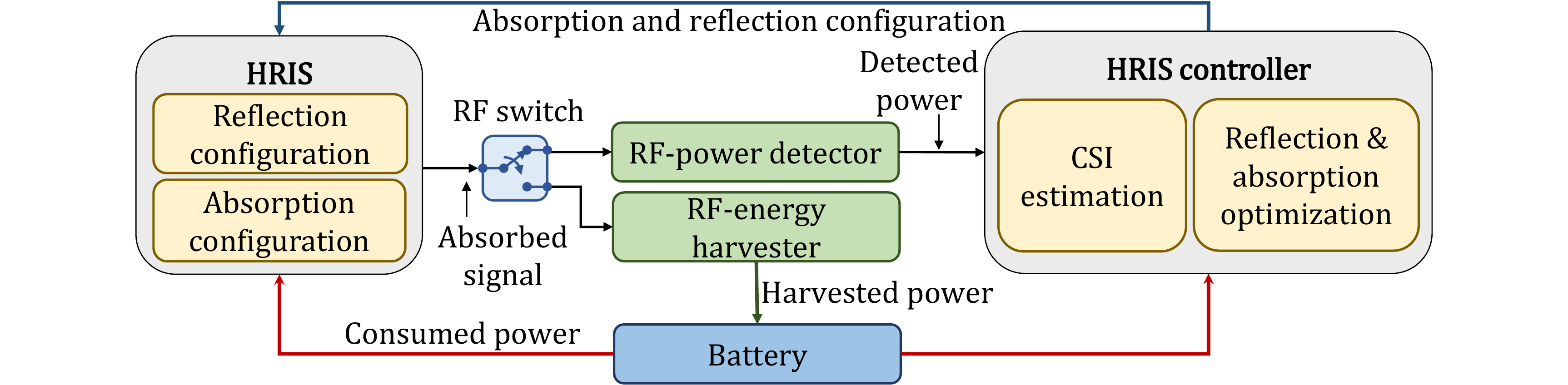}
        \caption{Overview of \name{}'s architecture and functional blocks}
        \label{fig:hris_block_diagram}
        \vspace{-0.1cm}
\end{figure}

\vspace{-0.2cm}
\section{\name{} overview}
\label{sec:framework_overview}

\begin{figure}[t]
        \center
        \includegraphics[width=1\linewidth]{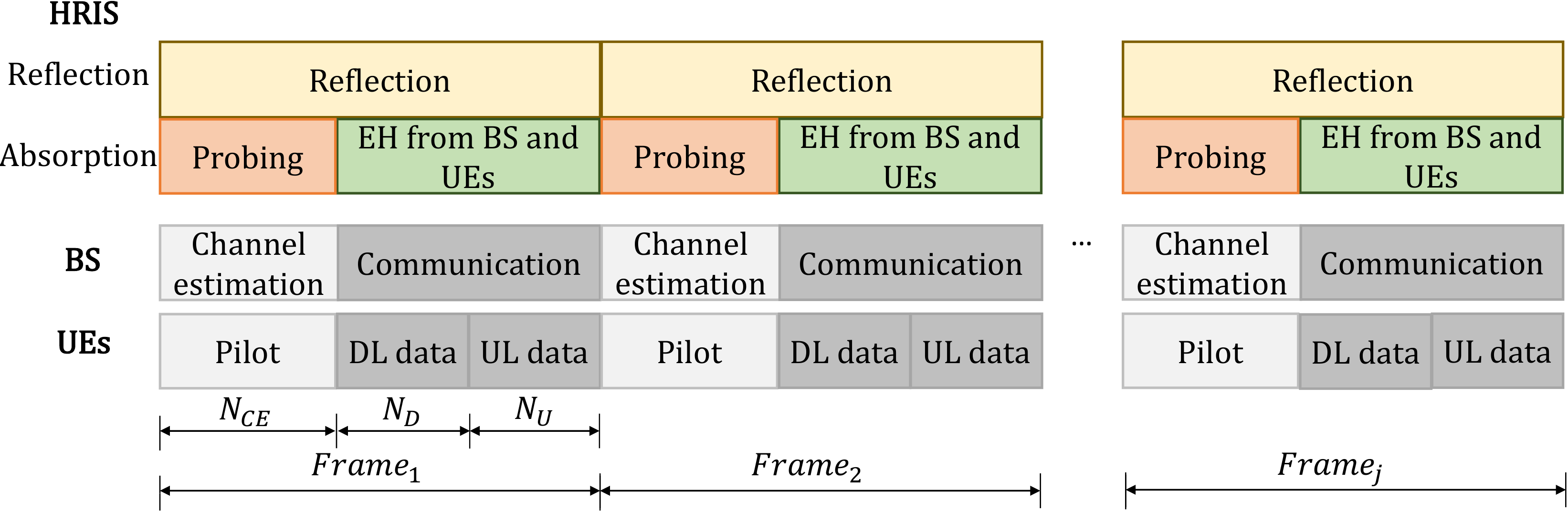}
        \caption{\label{fig:frame_structure} \name{} reference \gls{tdd} frame.}
        \vspace{-0.1cm}
\end{figure}

We depict the main building blocks of \name{} in Fig.~\ref{fig:hris_block_diagram} and remark that the reference \gls{hris} design has two branches, namely reflection and absorption. As the reflection branch is fully passive like in any typical \gls{ris}, here we highlight the path of the absorbed signal to the \gls{hris} controller, which outputs the \gls{hris} reflection and absorption configuration, as well as annotating the energy transfer to and from the service battery.     
To fully comprehend \name{} operation, we first introduce the reference communication frame. This is important to ensure that the autonomous operation of the \glspl{hris} is compliant with the standard by which \gls{bs} and \glspl{ue} communicate. 

\textbf{\name{} reference communication frame.} Without loss of generality, we assume that all devices use \gls{tdd} duplexing, specifically with the frame structure shown in Fig.~\ref{fig:frame_structure}. As described in Section~\ref{sec:hw_design}, the two \gls{hris} branches enable independent signal routes with different phase shift banks. This is implemented in the frame structure: while the \gls{hris} continuously reflects the impinging signals through the reflection branch, the absorption branch is involved in two fundamental operations, i.e. probing and \gls{eh}, and leverages on power-based indirect beamforming, which is described below. 
At the beginning of each frame, the \glspl{ue} transmit pilot signals over $N_{CE}$ slots while the \gls{bs} performs channel estimation to acquire end-to-end \gls{csi}\footnote{This frame structure follows \acrlong{mmimo} procedures~\cite{marzetta2015massive}.}. At the same time, such pilots are sensed by the \gls{hris} through probing to obtain a local estimate of the \gls{csi} (see Section~\ref{sec:hris_config}). It is worth highlighting that the end-to-end \gls{csi} would be affected by the additional high-gain paths provided by the \gls{hris} through the reflection branch. Hence, to minimize the channel estimation errors at the \gls{bs}, we assume that the \gls{csi} acquired at the \gls{hris} within a given communication frame is used to optimize the \gls{hris} reflection configuration in the subsequent one. Once the channel estimation phase is complete, the \gls{bs} optimizes the precoder and initiates the communication consisting of $N_U$ and $N_D$ slots dedicated to the \gls{ul} and the \gls{dl} direction, respectively. Concurrently, the \gls{hris} enables the energy harvester and selects the most suitable absorption configuration to harvest \gls{em} energy from the \gls{bs} (\gls{dl}) and the \glspl{ue} (\gls{ul}).

\textbf{Power-based indirect beamforming.} 
Both consumers of the absorbed signal in Figs.~\ref{fig:hris} and \ref{fig:hris_block_diagram}, i.e. the power detector and the energy harvester, require beamforming to enable channel estimation and \gls{eh}, respectively. 
Let us begin by identifying the challenges to be overcome when establishing autonomous \gls{hris} operations---probing and \gls{eh}---with only one \gls{rf} power detector, which can exclusively measure the sum-power of all the incident signals at every \gls{hris} meta-atom. Most available \gls{aoa} estimation techniques necessitate the signal samples at each receive antenna. Conversely, we make the most out of our proposed hardware design and perform an indirect estimation of the \gls{aoa}, by optimizing the phase shifts applied to the absorbed (sensed) signals at every meta-atom so as to maximize the power sensed by the detector. As adjusting the phase shifts applied by the meta-atoms is equivalent to realizing a virtual (passive) beamformer towards specified \glspl{aoa} of the incident signals with respect to the \gls{hris} surface, we can take advantage of the power sensing capability of the \glspl{hris} for estimating the \gls{bs}-\gls{hris} and \gls{hris}-\gls{ue} channels with very little local information, and accordingly configure the reflection and absorption branches to support communication and \gls{eh}.
}

\change{\textbf{\gls{hris} self-configuration.} The optimal configuration of the \gls{hris} is discussed in Section~\ref{sec:hris_config}. Here, we anticipate that any algorithmic solutions for optimizing the absorption and reflection properties of the \gls{hris} require the estimation of the \gls{csi} of the \gls{bs}-\gls{hris} and \gls{hris}-\gls{ue} channels. This results in a chicken-egg problem that needs to be tackled. To this end, in Section~\ref{sec:absorption_model}, we devise an online optimization approach that relies upon a finite set of \gls{hris} configurations, namely a codebook, that can be iteratively tested for probing a finite set of predefined \glspl{aoa}.}

\change{\textbf{\gls{hris} self-sufficiency.} 
The cheapest and most straightforward approach to enable \gls{hris} energy self-sufficiency without deploying an extensive electrical network is to endow each \gls{hris} with a dedicated battery, e.g. lithium-ion (Li-ion) or nickel metal hydride (NiMH). However, battery-operated \glspl{hris} call for regular maintenance, generating recurrent costs and additional complexity at \gls{ran} planning stage. Conversely, by introducing the ability to harvest the absorbed \gls{rf} energy, our proposed solution opens up the possibility of a self-sufficient \gls{hris} design that can sustain its operation by replenishing the service battery on the go. In Section~\ref{sec:energy_management}, we discuss in detail its feasibility in terms of required battery capacity and link its realization to a realistic evaluation of the \gls{hris} hardware consumption in Section~\ref{sec:hardware_consumption}.}

\vspace{-0.2cm}
\section{Model Design}
\label{sec:model_design}
In this section, we introduce the reference analytical model for the considered wireless \gls{hris}-aided network scenario. We start by stating the \gls{hris} configuration problem accounting for the practical limitations of such hardware platform.

\vspace{-0.3cm}
\subsection{Communication System Model}
\label{sec:sys_model}
We consider a scenario in which a \gls{bs} equipped with $M$ antennas serves $K$ single-antenna \glspl{ue} with the aid of an \gls{hris}. We model the \gls{bs} as a \gls{ula}, and the \gls{hris} as a \gls{pla} equipped with $N = N_x \times N_z$ meta-atoms, where $N_x$ and $N_z$ denote the number of elements along the $x$ and $z$ axis, respectively. We assume that the inter-distance of the \gls{bs} and \gls{hris} array elements is $\lambda/2$, where \change{$\lambda = c_0/f_c$ denotes the carrier wavelength, $f_c$ is the corresponding carrier frequency and $c_0$} is the speed of light. The joint reflection and absorption capabilities of the \gls{hris} are realized through directional couplers\footnote{A practical implementation of this architecture can be found in \cite{alexandropoulos2021hybrid}.} whose operation is determined by the parameter $\eta\in[0,1]$, which is the fraction of the received power that is reflected for communication, while $1-\eta$ is the amount of \change{absorbed power to be measured at the power detector or harvested at the energy harvester}.

We denote by $\vv{b}\in \mathbb{R}^3$, $\vv{r}\in \mathbb{R}^3$ and $\vv{u}_k \in \mathbb{R}^3$ the locations of the \gls{bs} center, the \gls{hris} center and the $k$-th \gls{ue}, respectively. Focusing on the \gls{dl}, the \gls{bs} transmits data to the $k$-th \gls{ue} over a direct \gls{los} link $\vv{h}_{\ssub{D},k} \in \mathbb{C}^{M\times 1}$ and a reflected link through the \gls{hris}. Such path can be decomposed into the \gls{los} channel $\vv{h}_k \in \mathbb{C}^{N \times 1}$ through which the \gls{hris} reflects the impinging signal towards the \gls{ue}, and the \gls{los} channel $\vv{G} \in \mathbb{C}^{N\times M}$ between the \gls{bs} and the \gls{hris}.

\change{
The array response vectors at the \gls{bs} and at the \gls{hris} towards the generic location $\vv{p} \in \mathbb{R}^3$ are denoted by $\vv{a}_{\ssub{BS}}(\vv{p}) \in \mathbb{C}^{M \times 1}$ and $\vv{a}_{\ssub{R}}(\vv{p}) \in \mathbb{C}^{N \times 1}$, respectively. Their elements are defined as 
    $\{\vv{a}_{\ssub{BS}}(\vv{p})\}_{m=1}^M \triangleq e^{j\langle\vv{k}_{\ssub{PB}}, (\vv{b}_m - \vv{b})\rangle}$, and $\{\vv{a}_{\ssub{R}}(\vv{p})\}_{n=1}^N \triangleq e^{j\langle\vv{k}_{\ssub{PR}} , (\vv{r}_n - \vv{r})\rangle}$,
where $\vv{k}_{\ssub{BP}}$ and $\vv{k}_{\ssub{PR}}$ are the wave vectors, which are defined as
\begin{equation}
    \vv{k}_{\ssub{PB}} \triangleq \frac{2\pi}{\lambda}\frac{\vv{p}-\vv{b}}{\|\vv{b} - \vv{p}\|}, \quad \text{and} \quad \vv{k}_{\ssub{PR}} \triangleq \frac{2\pi}{\lambda}\frac{\vv{p}-\vv{r}}{\|\vv{r} - \vv{p}\|},
\end{equation}
with $\vv{b}_m$ and $\vv{r}_n$ denoting the coordinates of the $m$-th \gls{bs} antenna element and of the $n$-th \gls{hris} meta-atom, respectively.}

The overall gain of a generic communication path between two given locations $\vv{p}$, $\vv{q} \in \mathbb{R}^3$ is defined as
$\gamma(\vv{p},\vv{q}) \triangleq \gamma_0 \left( d_0/\|\vv{p} - \vv{q}\| \right)^\chi$,
where $\gamma_0$ is the channel gain at a reference distance $d_0$ and $\chi$ is the pathloss exponent. Hence, the \gls{bs}-\gls{hris} and the \gls{hris}-$\text{\gls{ue}}_k$ channels are
\change{
\begin{align}
    \vv{G} &\triangleq \sqrt{\gamma(\vv{b},\vv{r})}\vv{a}_{\ssub{R}}(\vv{b})\vv{a}^{\herm}_{\ssub{BS}}(\vv{r}) \in \mathbb{C}^{N\times M}, \quad \text{and} \quad 
\end{align}
\begin{align}
    \vv{h}_k \triangleq \sqrt{\gamma(\vv{u}_k,\vv{r})}\vv{a}_{\ssub{R}}(\vv{u}_k) \in \mathbb{C}^{N\times 1}. \label{eq:channel_ris_ue}
\end{align}}
While, the direct \gls{bs}-$\text{\gls{ue}}_k$ channel is
\begin{equation}
    \vv{h}_{\ssub{D},k} \triangleq \sqrt{\gamma(\vv{b},\vv{u}_k)}\vv{a}_{\ssub{BS}}(\vv{u}_k) \in \mathbb{C}^{M\times 1}.
    \label{eq:channel_bs_ue}
\end{equation}
Thus, the received signal at the $k$-th \gls{ue} is 
\begin{equation}
y_k=\left(\sqrt{\eta}\vv{h}_k^{\herm}\vv{\Theta}\vv{G} +\vv{h}^{\herm}_{\ssub{D},k}\right)\vv{W} \vv{s} + n_k \in \mathbb{C},
\end{equation}
where $\vv{\Theta} = \mathrm{diag}[\alpha_1 e^{j\vartheta_1}, \dots, \alpha_N e^{j\vartheta_N}]$, with $\vartheta_i \in [0, 2\pi]$ and $|\alpha_i|^2 \leq 1$, $\forall i$ being the phase shifts and the gains introduced by the \gls{hris}, $\vv{W} \in \mathbb{C}^{M \times K}$ is the transmit precoding matrix whose $k$-th column $\vv{w}_k$ is the transmit precoder of $\text{\gls{ue}}_k$, $\vv{s} = [s_1, \dots, s_K]^\mathrm{T}$ is the transmit symbol vector with $\mathbb{E}[|s_k|^2] = 1$ $\forall k$, and $n_k$ is the noise term whose distribution is $\mathcal{CN}(0,\sigma_n^2)$. \change{Moreover, we assume that $\{\alpha_i\}_{i=1}^N$ and $\{\vartheta_i\}_{i=1}^N$ can be independently optimized.
Lastly, we underline that the phase shifters banks in Fig.~\ref{fig:hris} can apply different configurations to the incident signal on the reflection and absorption branches, respectively. 
This assumption allows decoupling the two \gls{hris} operations maximizing the harvested energy without affecting the communication task.}

\vspace{-0.2cm}
\subsection{\gls{hris} Optimization}
\label{sec:hris_config}

In this section, we focus on how an \gls{hris} can be endowed with self-configuring capabilities, and, in particular, how the absence of a dedicated control channel results in the need for the \gls{hris} of locally estimating the channels towards the \gls{bs} and the \gls{ue}, in order to establish and maintain a high-quality reflected path. To this end, we commence by formulating the optimization problem without imposing the absence of the control channel, and we then elaborate on the difficulty of solving the obtained problem by relying only on local \gls{csi} at the \gls{ris}.
The \gls{sinr} at the $k$-th \gls{ue} can be written as 
\begin{align}
     \!\!\!\!\!\!\mathrm{SINR}_k & = \frac{\big| \big(\sqrt{\eta}\vv{h}_k^{\herm} \vv{\Theta}\vv{G} + \vv{h}_{\ssub{D}.k}^{\herm})\vv{w}_k\big|^2}{\sigma^2_n + \sum_{j \neq k}\big|\big(\sqrt{\eta} \vv{h}_k^{\herm} \vv{\Theta}\vv{G} + \vv{h}_{\ssub{D}.k}^{\herm})\vv{w}_j \big|^2}, \label{eq:sinr_k}
\end{align}
where $\vv{w}_k$ is assumed to be given during the optimization of the configuration of the \gls{hris}. More precisely, $\vv{w}_k$ is optimized by the \gls{bs} after the channel estimation phase. The disjoint optimization of $\vv{w}_k$ and the \gls{ris} configuration facilitates the design and deployment of a control channel-free \gls{hris}. The optimization of $\vv{w}_k$ is elaborated in further text.
We aim at finding the optimal \gls{hris} configuration that maximizes the network sum-rate, which is directly related to the \gls{sinr} at every \gls{ue}.
More precisely, the network sum-rate is defined as $\mathrm{R} \triangleq \sum_{k=1}^K \log_2 \left(  1 \hspace{-0.05cm} + \mathrm{SINR}_{k} \right)$.

\change{
A feasible strategy to maximize the sum-rate is to optimize the \gls{hris} configuration so that the intensity of $\vv{h}_k^{\herm}\vv{\Theta}\vv{G}\vv{w}_k$ in \eqref{eq:sinr_k} is maximized, $\forall k$ i.e., the end-to-end \gls{ris}-assisted channel gain of each user is enhanced, which is approximately equivalent to maximizing the $\text{\gls{sinr}}_k$ as demonstrated in~\cite{albanese2022marisa}. Upon completion of this optimization, the \gls{bs} can optimize the precoding matrix $\vv{W}$. 
Indeed, even though the \gls{bs} cannot control the \gls{hris} configuration due to the absence of a control channel, it can always estimate the direct channel towards each \gls{ue} and the equivalent RIS-assisted link.
}

\textbf{Channel estimation and \gls{hris} configuration.} Let us now focus on the optimization of the \gls{hris} configuration, by taking into account that it is equipped with a single \gls{rf} power detector. To this end, we derive a closed-form expression for the \gls{hris} configuration that maximizes the reflected power.
We assume that a training phase exists, during which the \gls{bs} and each \gls{ue} transmit a pilot symbol $s$ in order to realize the initial beam alignment procedure\footnote{This standard procedure is essential before data transmission in, e.g., millimeter-wave networks for initial device discovery and channel estimation~\cite{ANR17_TWC}.}. Without loss of generality, we assume a certain degree of synchronization, i.e., the \gls{bs} and the \glspl{ue} transmit at different times, but all \glspl{ue} transmit simultaneously. We will relax this latter assumption in Section~\ref{sec:algorithm}. 

\change{For ease of presentation, we define $\vv{\uptheta} \in \mathbb{C}^{N \times 1}$ and $\vv{\upphi} \in \mathbb{C}^{N \times 1}$ representing the \gls{hris} configuration in the reflection and the absorption branch, respectively, as
\begin{align}
    \vv{\uptheta} &\triangleq [\alpha_1 e^{-j\vartheta_1}, \dots, \alpha_N e^{-j\vartheta_N}]^\mathrm{T} , \quad \text{and} \label{eq:phi}  \\
    \vv{\upphi} &\triangleq [\beta_1 e^{-j\varphi_1}, \dots, \beta_N e^{-j\varphi_N}]^\mathrm{T}, \nonumber
\end{align}
with $\vv{\Theta} = \mathrm{diag}(\vv{\uptheta}^{\herm})$.
The signals at the output of the \gls{rf} combiner on the absorption branch, which are obtained from the pilot signals transmitted by the \gls{bs} and the \glspl{ue}, are formulated as}
\change{
\begin{align}
    y_{\ssub{B}} & =   \sqrt{(1-\eta)} \, \vv{\upphi}^{\herm} \vv{G} \vv{w}_{\ssub{R}} s  + n \in \mathbb{C}, \quad \text{and} \label{eq:y} \\
    y_{\ssub{U}}  &=  \sqrt{(1-\eta)} \, \vv{\upphi}^{\herm} \vv{h}_{\ssub{\Sigma}} s  + n \in \mathbb{C}, \nonumber
\end{align}}
where we assume that the \gls{bs} and the \glspl{ue} emit the same amount of power $P$ and $n \sim \mathcal{N}(0,\sigma_n^2)$ is the additive noise term. 
Let $\vv{w}_{\ssub{R}}$ be the optimal \gls{bs} precoder for the \gls{bs}-\gls{hris} link. We will see shortly that the knowledge of $\vv{w}_{\ssub{R}}$ is not explicitly needed to optimize the \gls{hris} configuration. Also, we define $\vv{h}_{\ssub{\Sigma}} \triangleq \sum_{k=1}^K \vv{h}_k$. Since the \gls{ue}-\gls{hris} channel $\vv{h}_k$ corresponds to the \gls{ul}, to use it in the \gls{dl}, we assume that the channel reciprocity holds.

\change{Therefore, the power $P_{\ssub{B}}$ and $P_{\ssub{U}}$ available for the power detector or the energy harvester downstream of the \gls{rf} combining circuitry can be formulated as the expectation $\Exp \left[| y_b|^2\right]$ and $\Exp \left[| y_u|^2\right]$, which is}
\begin{align}
    P_{\ssub{B}} & = (1-\eta) \, \big|\vv{\upphi}^{\herm}\vv{G} \vv{w}_{\ssub{R}} s\big|^2 + \sigma_n^2, \quad \text{and} \\
    P_{\ssub{U}} & =  (1-\eta) \, \big|\vv{\upphi}^{\herm} \vv{h}_{\ssub{\Sigma}}s\big|^2 + \sigma_n^2 \label{eq:p_UE_BS},
\end{align}
\change{when absorbed from the pilot signals emitted by the \gls{bs} and the \glspl{ue}, respectively.}
In order to be self-configuring, an \gls{hris} needs to infer the channels $\vv{G}$ and $\vv{h}_\ssub{\Sigma}$ only based on $P_{\ssub{B}}$ and $P_{\ssub{U}}$ in \eqref{eq:p_UE_BS}, respectively.
This is equivalent to finding the configuration of the \gls{hris} that maximizes $P_{\ssub{B}}$ and $P_{\ssub{U}}$, which in turn corresponds to estimating the directions of incidence of the signals on the \gls{hris}.
As a result, we formulate the following optimization problem, whose solution is the \gls{hris} configuration that maximizes $P_{\ssub{B}}$
\begin{align}
    \max_{\vv{\upphi}} & \ \ |\vv{\upphi}^{\herm}\vv{G}\vv{w}_R|^2 \label{eq:obj_hris_conf_BS} \\
    \textup{s.t.} & \ \ |{\phi}_i|^2\leq 1 \quad i = 1,\dots,N, \nonumber
    \label{eq:hris_conf_BS}
\end{align}
where ${\phi}_i$ is the $i$th element of $\vv{\upphi}$. The objective function in~\eqref{eq:obj_hris_conf_BS} can be recast as
\begin{align}
    |\vv{\upphi}^{\herm}\vv{G}\vv{w}_R|^2 = \vv{\upphi}^{\herm}\vv{a}_{\ssub{R}}(\vv{b})\vv{a}_{\ssub{R}}^{\herm}(\vv{b})\vv{\upphi} \, \large|z_{\ssub{R,R}}\large|^2,
\end{align}
where $z_{\ssub{R,R}} \triangleq \sqrt{\gamma(\vv{b},\vv{r})}\vv{a}_{\ssub{BS}}^{\herm}(\vv{r})\vv{w}_\ssub{R} \in \mathbb{C}$ is the projection of the \gls{bs} precoding vector $\vv{w}_{\ssub{R}}$ onto the \gls{bs}-\gls{hris} direction. 
Hence, the optimal \gls{hris} configuration for maximizing the absorbed power from the \gls{bs} is $\vv{\upphi}_{\ssub{B}} \in \mathbb{C}^{N \times 1}$ with
\begin{align}
\phi_{\ssub{B},i} = e^{j \angle a_{\ssub{R},i}(\vv{b})}, \quad i = 1,\dots,N. \label{eq:v_B}
\end{align}

Analogously, the optimal \gls{hris} configuration that maximizes $P_{\ssub{U}}$ is $\vv{\upphi}_{\ssub{U}} \in \mathbb{C}^{N \times 1}$ with
\begin{align}
\phi_{\ssub{U},i} = e^{j \angle h_{\ssub{\Sigma},i}}, \quad i = 1,\dots,N. \label{eq:v_U}
\end{align}

From \eqref{eq:v_B} and \eqref{eq:v_U}, we evince, as anticipated, that the optimal \gls{hris} configuration that maximizes the sensed power \change{(i.e., the \gls{hris} absorption configuration)} depends only on the \gls{hris} array response vectors towards the \gls{bs} and \gls{ue} directions, but it is independent of the (optimal) \gls{bs} precoding.

Based on $\vv{\upphi}_{\ssub{B}}$ and $\vv{\upphi}_{\ssub{U}}$, we are now in the position of proposing a distributed approach for optimizing the \gls{hris} \change{configuration in the reflection branch $\vv{\uptheta}$ (i.e., the \gls{hris} reflection configuration)}. In particular, we formulate the following optimization problem.
\begin{problem}[Multi-\gls{ue} \gls{sinr}-based \gls{hris} configuration]\label{problem:max_sinr_multi_ue} 
\begin{align}
    \max_{\vv{\Theta}} & \ \ \frac{\big|\vv{h}_{\ssub{\Sigma}}^{\herm} \vv{\Theta} \vv{G}  \vv{w}\big|^2}{\sigma_n^2}    \label{eq:max_sinr_multi_ue}\\
   \textup{s.t.} & \ \  |\Theta_{ii}|^2 \leq  1 \quad i = 1,\dots, N. \nonumber 
\end{align}
\end{problem}
\change{
Problem \ref{problem:max_sinr_multi_ue} is independent of the direct channels between the \gls{bs} and the \glspl{ue}, as well as of the \gls{bs} precoder $\vv{w}$: these are fundamental requirements due to the lack of control channel. Notably, the objective function in \eqref{eq:max_sinr_multi_ue} is a lower bound for the sum of the powers of the signals transmitted by the \glspl{ue} independently, i.e. $\big|\vv{h}_{\ssub{\Sigma}}^{\herm} \vv{\Theta} \vv{G}  \vv{w}\big|^2 \leq \sum_{k=1}^K \big|\vv{h}_{k}^{\herm} \vv{\Theta} \vv{G}  \vv{w}\big|^2$, which are reflected by the \gls{hris}. This allows us to reformulate the objective function in~\eqref{eq:max_sinr_multi_ue} as}
\begin{equation}
    \frac{\big|z_{\ssub{R}}\vv{\uptheta}^{\herm}\hat{\vv{h}}\big|^2}{\sigma_n^2}, \label{eq:sinr_multi_ue_2}
\end{equation}
where $\hat{\vv{h}} \triangleq \vv{h}^*_{\ssub{\Sigma}} \circ \vv{a}_\ssub{R}(\vv{b})$ is the  equivalent channel that accounts for the overall effect of the aggregate \gls{ue}-\gls{hris} channels from the \gls{hris} standpoint, and $z_{\ssub{R}} \triangleq \sqrt{\gamma(\vv{b},\vv{r})}\vv{a}_{\ssub{BS}}^{\herm}(\vv{r})\vv{w} \in \mathbb{C}$ is the reflected path between the \gls{bs} and the \gls{hris} for a given precoder $\vv{w}$ at the \gls{bs}. 

Therefore, the \gls{hris} optimal configuration solution of Problem \ref{problem:max_sinr_multi_ue} is
\begin{align}
    \vv{\uptheta}_{\ssub{BU}}=e^{j\angle \  \hat{\vv{h}}}=e^{j\angle \left(\vv{h}_{\ssub{\Sigma}}^* \ \circ \ \vv{a}_{\ssub{R}}(\vv{b})\right)} = \vv{\upphi}_{\ssub{U}}^* \circ \vv{\upphi}_{\ssub{B}}, \label{eq:v_BU}
\end{align}
which proves that the \gls{hris} configuration in the absence of a control channel can be inferred solely from $\vv{\upphi}_{\ssub{U}}$ and $\vv{\upphi}_{\ssub{U}}$.

\change{
\vspace{-0.2cm}
\subsection{Energy Management Model}
\label{sec:energy_management}

In the following, we extend the above-mentioned \gls{hris} self-configuration capabilities and propose our energy self-sufficient design. It is worth highlighting that the \gls{hris} does not perform self-configuration and energy harvesting concurrently.
In particular, as described in Section~\ref{sec:hris_config}, the \gls{hris} self-configuration requires measuring the sensed power of pilot signals transmitted by the \gls{bs} and the \glspl{ue} during the probing phase. On the other hand, once configured, the \gls{hris} harvests energy from the \gls{rf} signal emitted by the \gls{bs} and the \glspl{ue} while communicating, i.e. from transmissions over the \gls{dl} \gls{bs}-\gls{hris} and the \gls{ul} \glspl{ue}-\gls{hris} links\footnote{\change{We assume the \gls{dl} and \gls{ul} wireless channels to be reciprocal.}}.
When performing \gls{eh}, the energy harvester converts the absorbed \gls{rf} power to electrical power, which we use to operate the \gls{hris} without the need for an external power supply. Moving forward, we characterize both the energy collected and consumed by the \gls{hris}, and identify the fundamental design and operational choices behind the implementation of an \gls{eh}-aided self-sufficient \gls{hris}.

\textbf{Energy harvesting and consumption.} Given the configuration $\vv{\upphi}$ in \eqref{eq:phi} of the \gls{hris} absorption branch (i.e., the \gls{hris} absorption configuration) and the input \gls{rf} power $P_B$ and $P_U$ in \eqref{eq:p_UE_BS} to the energy harvester, the harvested power from the \gls{bs}-\gls{hris} and \glspl{ue}-\gls{hris} transmissions can be respectively expressed as
\begin{equation}
    P_{H,B} = f(P_B), \quad \text{and} \quad P_{H,U} = f(P_U),
\end{equation}
with $f(\cdot)$ denoting a non-linear function representing the \gls{in-out} characteristic of the harvester, namely
\begin{equation}
    f(x) = \frac{ax + b}{x + c}-\frac{b}{c}, 
    \label{eq:harvested_power}
\end{equation}
where $a$, $b$, and $c$ are constants fitting the \gls{in-out} response of the harvester~\cite{Chen2017}.
We consider the \gls{dl} and \gls{ul} \glspl{ue}-\gls{bs} transmissions to be scheduled according to a typical \gls{tdd} frame consisting of $N_D$ \gls{dl} and $N_{U}$ \gls{ul} transmission time slots, as sketched in Fig.~\ref{fig:frame_structure}. Hence, the overall harvested power is obtained as 
\begin{equation}
    P_H =\xi \left(N_D \, P_{H,B} + N_U \, P_{H,U} \right),
    \label{eq:harvested_energy_2}
\end{equation}
where we introduce the factor $\xi \in [0,1]$ to model that $i$) each time slot corresponds to a transmission opportunity for the \gls{bs} and the \glspl{ue}, which is taken according to the level of traffic to be transmitted by each device; $ii$) the \gls{eh} is performed alternatively to the self-configuration function, thus for a fraction of the total active time, as to keep the \gls{hris} configuration up-to-date while collecting enough \gls{rf} power to operate.

We now analyze the \gls{hris} power consumption. While from the communication standpoint, the \gls{hris} is a passive device, our design requires (minimal) electrical power to operate its active components, namely the power detector when measuring the absorbed \gls{rf} power and the meta-atoms to hold the desired phase shifts configuration.
As the power consumption depends on the specific \gls{hris} hardware, it can be modelled as
\begin{equation}
    P_C = g(\vv{\uptheta},\vv{\upphi}, \text{hardware}),
    \label{eq:energy_consumption_general}
\end{equation}
where $g(\cdot)$ is a custom function taking into account the instantaneous \gls{hris} reflection and absorption configuration for a specific hardware design, which is tailored to a realistic \gls{hris} implementation in Section~\ref{sec:hardware_consumption}.  Let us consider the alternation between \gls{eh} and power detection (i.e., self-configuration) being periodic every $T$ second, i.e. reconfiguration period. The average harvested and consumed energy in each reconfiguration period can be written as
\begin{equation}
    E_H = T P_H, \quad \text{and} \quad E_C = T P_C.
\end{equation}

\begin{figure}[t]
        \center
        \includegraphics[width=1\linewidth]{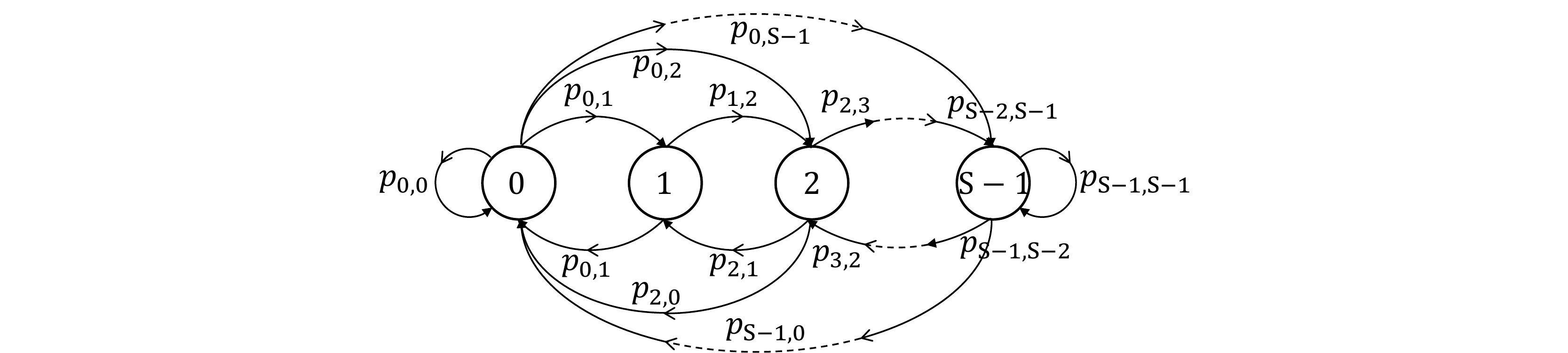}
        \caption{\label{fig:markov_chain} Reference Markov-chain-based battery state transition diagram.}
        \vspace{-0.1cm}
\end{figure}

\textbf{Markov-chain-based energy storage model.} The derivation of the conditions for energy self-sufficiency requires a model for the \gls{hris} battery. Specifically, we represent the energy storage in the \gls{hris} battery as a homogeneous Markov process\footnote{\change{\Gls{mc} models for energy storage have been empirically proven in the literature, see e.g.~\cite{Li2008markov}.}}, in which the next battery \gls{soc} depends only on the current one, satisfying the so-called Markov \textit{memorylessness} property~\cite{song2012development}. 
Let us consider the \gls{mc} model in Fig.~\ref{fig:markov_chain}, whose $S \in \mathbb{N}\setminus\{0,1\}$ discrete states correspond to the possible \gls{hris} battery states of charge, spanning from state $0$, i.e., fully discharged, to state $S-1$, i.e., fully charged. Without loss of generality, we consider the difference in energy between two adjacent states $\Delta \in \mathbb{R}$ to be constant. Thus, we can define the battery capacity as
\begin{equation}
    C = (S-1) \,  \Delta,  \label{eq:capacity}
\end{equation}
under the assumption of linear charge and discharge processes. We indicate the transition probability matrix (or stochastic matrix) as $\Psi \in [0,1]^{S \times S}$, wherein each element $p_{i,j}$ denotes the probability of moving from state $i$ to state $j$. The amount of energy stored in the battery in each time unit is not constant, thereby leading to possible multiple state changes. In fact, we have no prior knowledge on the amount of harvested energy from the \gls{bs} and \glspl{ue} during their communication, as it is tightly coupled to several external factors as per \eqref{eq:harvested_energy_2}. Therefore, for ease of derivation, we consider the net stored energy per unit of time $\Delta E = E_H - E_C$ to be randomly distributed with \gls{cdf} $F_{\Delta E}$.  
Besides, we assume the associated stochastic process to be stationary as to guarantee that its unconditional probability distribution remains constant when shifted in time (i.e., for every unit of time).   

The transition probabilities can be written as
\begin{align}
p_{0,0} &= P[\Delta E < \Delta],\\
p_{i,j} &= P[(j-i) \Delta \leq \Delta E < (j-i+1) \Delta], \\
& 0<i<S-1, \, 0<j<S-1, \nonumber \\
p_{S-1,S-1} &= P[\Delta E > - \Delta],
\end{align}
where we reflect that a net stored energy $\Delta E \geq \Delta$ is required to move up from the current state, while $\Delta E < 0$ leads to a transition to a previous state. For instance, transitioning from the current \gls{soc} to the following one demands that $\Delta \leq \Delta E < 2 \Delta$, while a two-state transition upwards requires that $2\Delta \leq \Delta E < 3 \Delta$. Furthermore, additional depletion of charge from the battery in state $0$ or increase of charge in state $S-1$ do not change the current \gls{mc} state, i.e. the extremes of the \gls{soc} intervals are saturation points due to the underlying physical implementation of the battery management circuitry.   

As mentioned above, $\Delta E$ is a continuous \gls{rv}, which allows deriving an equivalent and more compact formulation of the transition probabilities by means of its \gls{cdf} $F_{\Delta E}$, namely $P[\Delta E \leq x] = F_{\Delta E} (x)$.
Hence, the transition probabilities can be re-written as
\begin{align}
p_{0,0} &= F_{\Delta E} (\Delta),\\
p_{i,j} &= F_{\Delta E} ((j-i+1)\, \Delta) - F_{\Delta E} ((j-i)\, \Delta), \\ & 0<i<S-1, \, 0<j<S-1, \nonumber \\
p_{S-1,S-1} &= 1- F_{\Delta E} (\Delta).
\end{align}

In the steady state, the limiting probabilities are obtained as $\vv{\pi} = \Psi^{\tran} \vv{\pi}$, where $\vv{\pi} \in [0,1]^{S \times 1}$ is the stationary distribution indicating the expected probability of ending in each of the states upon the convergence of the \gls{mc}. It is worth highlighting that we prevent state $0$ from being an absorbing state for the \gls{mc}, namely a state that once entered cannot be left, by assuming that the \gls{hris} with {a low battery \gls{soc}} defaults to an idle configuration that does not consume power while enabling power absorption, just at a lower efficiency. In particular, to achieve negligible power consumption,
we deactivate all phase shifters when the battery \gls{soc} falls below a minimum guard threshold $\Gamma$, thereby making the \gls{hris} only capable of harvesting power from signals incoming from within the solid angle enhanced by the idle beamforming configuration. This results in a loss of efficiency, which under the assumption of uniformly distributed \glspl{ue} in the service area, returns a fraction of the harvested power, namely $P_{H,\Gamma} = \nu P_H$,
where $\nu = \nu_x \, \nu_y = B_x\, B_y/\pi^2$, with $B_x$, $B_y$ denoting the beamwidth on the horizontal and vertical axis of the corresponding beamformer. Following~\cite{albanese2022ris}, we consider $B_x=1/(N_x\delta)$, $B_y=1/(N_y\delta)$, whereas $\delta = d/\lambda$ is the \gls{hris} element-spacing wavelength ratio.

Given the above formulation, we are in the position to analyze the energy self-sufficiency condition for the \gls{hris}{, i.e., the condition in which the \gls{soc} is greater than $\Gamma$}.
{We define as $p_{LoC}$ the probability of \gls{loc}, namely the probability of the \gls{mc} not being able to meet a negative net stored energy {without falling below $\Gamma$} given its current \gls{soc}.
In other words, 
\begin{equation}
    p_{LoC} = \sum_{j = 0}^{S_{\Gamma}} \pi_j,
    \label{eq:p_loc}
\end{equation}
where we consider all the possible mutually exclusive events leading to a \gls{loc}, and where we denote as $S_{\Gamma}$ the state of the \gls{mc} corresponding to the minimum \gls{soc} guard threshold.}
As per \eqref{eq:p_loc}, $p_{LoC}$ depends on the stationary probabilities $\vv{\pi}$ and the transition probability matrix $\Psi$, which in turn are affected by the number of states $S$, the energy difference between subsequent states $\Delta$, and the statistics of the net stored energy $\Delta E$. 
Therefore, for a given $\Delta E$ distribution, we perform a linear search on $S$ and a set of feasible $\Delta$'s and derive $\vv{\pi}$ and $\Psi$ to approach the design $p_{LoC}$. The corresponding minimal battery capacity is then given by \eqref{eq:capacity}. 
}

\vspace{-0.2cm}
\section{Codebook-Based Optimization of \glspl{hris}}
\label{sec:absorption_model}

\change{Problem~\ref{problem:max_sinr_multi_ue} provides us with a mathematical model for optimizing the \gls{hris} configuration based on information gathered on the absorption branch. From a practical standpoint, however, the optimal solution $\vv{\uptheta}_{\ssub{BU}}$ in \eqref{eq:v_BU} depends on the array response vectors from the \gls{hris} towards the \gls{bs} and the \glspl{ue}. To implement the obtained solution, the array response vectors, i.e., the \gls{bs}-\gls{ris} and \gls{ris}-\glspl{ue} \glspl{aoa}, need to be estimated, but this is not possible at the \gls{hris} due to the absence of \gls{rf} chains and of a control channel. In this section, we propose a codebook-based approach for estimating the necessary \glspl{aoa} and then computing  $\vv{\uptheta}_{\ssub{BU}}$, $\vv{\upphi}_{\ssub{B}}$ and $\vv{\upphi}_{\ssub{U}}$ in a distributed manner and locally at the \gls{hris}, i.e., our proposed \name{}.}

\vspace{-0.2cm}
\subsection{\name{}}\label{sec:algorithm}
\name{} optimizes the \glspl{hris} based on an \change{appropriately designed codebook~\cite{albanese2022marisa}}
, which allows for the estimation of the \gls{bs}-\gls{ris} and \gls{ris}-\glspl{ue} \glspl{aoa} in a distributed manner. The use of codebooks is a known approach in \gls{ris}-assisted communications, e.g., \cite{WZ2020_TCOM,He2020}, and it is usually implemented by assuming that the electronic circuits of the \gls{ris} can realize a finite number of phase responses (e.g., through PIN diodes~\cite{boles2011algaas}). Therefore, our proposed \name{} is compatible with conventional implementations of \glspl{ris}, but it does not need a control channel.

Let us consider a codebook $\mathcal{C} = \{ \vv{c}_1,\dots,\vv{c}_\ssub{L}\}$, whose codewords $\vv{c}_l \in \mathbb{C}^{N\times 1}$ are unit-norm beamforming vectors that correspond to a discrete set of possible phase shift matrices $\vv{\Theta}_l = \mathrm{diag}(\vv{c}_l^{\herm})$. In particular, each codeword $\vv{c}_l$ is constituted by discrete-valued entries that mimic a sort of phase quantization. The discrete values of the codewords are assumed to belong to the following set 
\begin{equation}
    \mathcal{Q} = \bigg\{\frac{2\pi}{2^{Q}}m : m = 0,\dots, 2^{Q-1}, m \in \mathbb{N} \bigg\}, \label{eq:phase}
\end{equation}   
where $2^Q$ is the possible number of discrete values.

\change{
In \name{}, the \gls{hris} operates in a \gls{tdd} fashion, as introduced earlier in Section.~\ref{sec:framework_overview}, which includes \textit{probing} and \gls{eh} (absorption branch), and \textit{reflection} (reflection branch). During probing, the \gls{hris} estimates the \glspl{aoa} that correspond to the \gls{bs} and to the \glspl{ue}. Upon probing completion, the \gls{hris} has gathered enough information to derive and set the reflection and absorption configuration as to assists the transmission of data between the \gls{bs} and the \glspl{ue} while harvesting enough \gls{em} energy to sustain its own operation.   

}

Without loss of generality, we assume that each codeword of the codebook is, to a certain extent, spatially directive, i.e., the resulting \gls{hris} configuration maximizes the absorbed power only in correspondence of a (narrow) solid angle. This is relatively simple to realize by enforcing, e.g., some constraints on the design of the condewords in terms of half-power beamwidth of the corresponding radiation pattern of the \gls{hris}.
Therefore, by iteratively sweeping across all the codewords $\vv{c}_l \in \mathcal{C}$, the \gls{hris} can scan, with a given spatial resolution, the \gls{3d} space and can detect network devices (the \gls{bs} and the \glspl{ue}) by using pilot signals emitted only by those devices. During this probing phase, the \gls{hris} collects a set of power measurements, or equivalently a power profile, $\mathcal{P} = \{\rho_1,\dots,\rho_L\}$ where each element $\rho_l \in \mathbb{R}$ is the power level sensed (measured) by the \gls{hris} when using the codeword $\vv{c}_l$. As a result, the array response vectors in $\vv{G}$ or $\vv{h}_{\Sigma}$ can be estimated from $\mathcal{P}$. In practice, this boils down to detecting the peaks in $\mathcal{P}$ and identifying the corresponding angular directions. By construction, in fact, the \gls{hris} detects a power peak only if there is at least one transmitter in the direction synthesized by the  \gls{hris} beampattern, i.e. the considered codeword. The finer the angular selectivity of the \gls{hris}, the longer the probing phase. Therefore a suitable compromise needs to be considered. In particular, we assume that $\rho_l$ is a power peak in $\mathcal{P}$ if it is greater than a given threshold $\tau \in \mathbb{R}^+$. Let $\mathcal{I} \triangleq \{i < L \ : \ \rho_i \in \mathcal{P} > \tau \}$ be the set of indexes $l$ corresponding to the power peaks.
Then, depending on which devices transmit their pilot signals, $\vv{\upphi}_\ssub{B}$ and $\vv{\upphi}_{\ssub{U}}$ in \eqref{eq:v_BU} can be estimated as
\begin{equation}
    \vv{\upphi}_\ssub{B} = \sum\nolimits_{i \in \mathcal{I}} \delta_i \vv{c}_i, \quad \quad \vv{\upphi}_{\ssub{U}} = \sum\nolimits_{i \in \mathcal{I}} \delta_i \vv{c}_i,
    \label{eq:v_B_code}
\end{equation} 
where $\delta_i \in \{1, \rho_i \}$ is a weight parameter that allows performing a hard ($\delta_i=1$) or a soft ($\delta_i=\rho_i$) combining of the power peaks in $\mathcal{I}$ based on the actual measured power $\rho_i$. \change{Therefore, probing directly returns the \gls{hris} absorption configurations $\vv{\upphi}_\ssub{B}$ and $\vv{\upphi}_{\ssub{U}}$ to perform \gls{eh} from the \gls{bs} and the \glspl{ue}, respectively.}
The end-to-end \gls{hris} optimal configuration $\vv{\uptheta}_{\ssub{BU}}$ for reflection is first computed from \eqref{eq:v_BU} and is then projected onto the feasible set of discrete phase shifts in \eqref{eq:phase}, which eventually yields the desired  $\bar{\vv{\uptheta}}_{\ssub{BU}}$. The proposed probing phase is summarized in Algorithm~\ref{alg:marisa_prob}.

\begin{algorithm}[t!]
\scriptsize
  \caption{\name{} -- Probing phase}\label{alg:marisa_prob}
  \begin{algorithmic}[1]
     \State Data: $\mathcal{C}$, $\tau \in \mathbb{R}^+$ 
     \State Perform a beam sweeping setting $\vv{\Theta}_l = \mathrm{diag}(\vv{c}_l)$, $\forall 
     \vv{c}_l \in \mathcal{C}$
     \State Measure the corresponding power profile $\mathcal{P}$
    \State Obtain $\mathcal{I} = \{i < L \ : \ \rho_i \in \mathcal{P} > \tau \}$
    \If{the \gls{bs} transmits the pilot signals}
    \State \change{Compute $\vv{\upphi}_{\ssub{B}} = \sum_{i \in \mathcal{I}} \delta_i \vv{c}_i^{\herm}$}
    \ElsIf{the \glspl{ue} transmit the pilot signals}
    \State \change{Compute $\vv{\upphi}_{\ssub{U}} = \sum_{i \in \mathcal{I}} \delta_i \vv{c}_i^{\herm}$}
    \EndIf
    \State \change{Obtain $\vv{\uptheta}_{\ssub{BU}} = \vv{\upphi}_{\ssub{B}} \circ \vv{\upphi}_{\ssub{U}}^*$}
    \State \change{Obtain $\bar{\vv{\uptheta}}_{\ssub{BU}}$ (reflection), $\bar{\vv{\uptheta}}_{\ssub{B}}$ and $\bar{\vv{\uptheta}}_{\ssub{B}}$ (absorption) after quantization}
    \end{algorithmic}
\end{algorithm}

\change{\textbf{Effect of dependent phase shifters.}
It is worth pointing out that if the implementation of the \gls{hris} hardware relies on dependent (instead of independent, see Section~\ref{sec:hw_design}) phase shifters banks, i.e., the configurations of the reflection and the absorption branches are coupled, sub-optimal configurations (optimized for power detection) of the \gls{hris} would be iteratively set during the probing phase, leading to a reduction of the communication performance as they would concurrently generate reflections in unwanted directions. Likewise, another compromise would arise when performing energy harvesting during communication, being the two functions fundamentally different in the way they affect the \gls{rf} power flow, i.e., enhancing and suppressing the reflected path, respectively. For the sake of brevity, we omit a deeper analysis of the performance trade-offs of such simpler hardware design and henceforth only consider a design based on independent phase shifters.      
}

\change{
\vspace{-0.7cm}
\subsection{Configuration-related hardware consumption}
\label{sec:hardware_consumption}
\vspace{-0.2cm}

As mentioned in Section~\ref{sec:energy_management}, the \gls{hris} power consumption heavily depends on its actual hardware implementation. In the following, we analyze the consumed power $P_C$, tailoring its general definition formulated in ~\eqref{eq:energy_consumption_general} to the hardware design proposed in Section~\ref{sec:hw_design}.  

We commence by noticing that a consumption model for \glspl{hris} leveraging meta-atoms is not yet established in the literature. Therefore, for the sake of tractability, we assimilate each meta-atom to a set of passive components interconnected by a variable number of \gls{pin} diodes~\cite{petrou2022first, dai2020reconfigurable}. \gls{pin} diodes can be activated (ON state) or deactivated (OFF state) to control the \gls{em} properties of the meta-atom and modulate the configuration of the related phase shift~\cite{dai2020reconfigurable}.
Each meta-atom can assume $\mathcal{Q}$ different possible phase configuration values, each corresponding to a different combination of activation of the \gls{pin} diodes forming the atom. Therefore, without loss of generality,
we consider a number of \gls{pin} diodes equal to the quantization level $Q$ in \eqref{eq:phase}. Moreover, we assume a direct mapping between configuration index $m$, corresponding to the $m$-th phase configuration in $\mathcal{Q}$, and the corresponding activation pattern of the \gls{pin} diodes, which we consider to be equal to the binary counterpart $m_2$ of the index $m$.
We can then model the power absorbed by the $m$-th phase configuration as the cumulative power absorbed by all the active diodes, which can be expressed as
\begin{equation}
    P_{atom}(m) = P_{ON} \left(m - \sum_{i=1}^{Q} \left\lfloor \frac{m}{2^i} \right\rfloor \right),
    \label{eq:meta_atom_power}
\end{equation}
wherein $P_{ON}$ denotes the power absorbed by a single \gls{pin} diode when set to the ON state.
Given the \gls{hris} configuration $\vv{c}$ belonging to the feasible set $\mathcal{Q}$, the overall absorbed power is
\begin{equation}
    P_C(\vv{c}) = \sum_{n=1}^{N} P_{atom}(\kappa_{\vv{c},n}),
    \label{eq:codeword_power}
\end{equation}
where $\kappa_{\vv{c},n} \in \{ 0, ..., 2^{Q-1} \}$ is the activation pattern of the \gls{pin} diodes of the $n$-th element of the \gls{hris} when $\vv{c}$ is set. Interestingly, the absorbed power depends on the configuration itself, hinting at the possibility of jointly optimizing the communication and energy consumption properties of the codebook, which however is out of the scope of this paper. Nonetheless, we refer the reader to~\cite{albanese2022marisa} for an example of codebook optimization without energy consumption considerations.
}

\vspace{-0.2cm}
\section{Performance Evaluation}
\label{sec:results}
To prove the feasibility of \name{}, we evaluate it in different scenarios and compare it against the \gls{soa} benchmark scheme, recently reported in~\cite{Mursia2021}, which relies upon a control channel to perform a centralized optimization.
The simulation setup and the parameters are given in Table~\ref{tab:parameters}. All results are averaged over $100$ simulation instances. 

\begin{table*}[t!]
\vspace{-0.2cm}\caption{Simulation setup and parameters}
\label{tab:parameters}
\centering
\resizebox{\linewidth}{!}{%
\begin{tabular}{c|c|c|c|c|c|c|c|c|c|c|c|c}
\rowcolor[HTML]{EFEFEF}
\textbf{Parameter}  & $P$      & $M$ & $N_x,N_z$ & $f_c$    & $\vv{b}$       & $\vv{r}$                               & $A$                  & $\zeta$         & $\lambda_{\ssub{B}}$ & $h_{\ssub{B}}$  &  $r_{\ssub{B}}$ & $N_{D},N_{U}$\\  
\textbf{Value}      & $20$ dBm & $4$ & $8,4$     & $28$ GHz & $(-25,25,6)$ m & $(0,0,6)$ m                            & $50 \times 50$ m$^2$ & $0.5$           & $0.3$ m$^{-2}$       & $1.8$ m         &  $0.6$ m        & $8,3$\\
\hline
\rowcolor[HTML]{EFEFEF}
\textbf{Parameter}  & $L$      & $Q$ & $\eta$    & $T$      & $P_{ON}$       & $\chi_{\text{LoS}},\chi_{\text{NLoS}}$ & $\sigma^2_n$         & $d_0, \gamma_0$ & $C$                  & $\Delta$        & $\Gamma$        & $T$\\
\textbf{Value}      & $32$     & $2$ & $0.8$     & $10$ ms  & $0.10$ mW      & $2$, $4$                               & $-80$ dBm            & $1$             & $400$ mAh            & $20$ mAh        & $10\%$          & $10$ ms\\
\end{tabular}%
}
\vspace{-0.2cm}
\end{table*}

The network area $A$ is a square, and the \gls{bs} and the \gls{hris} (or the \gls{ris}) are located in the midpoints of two of its adjacent edges. The \glspl{ue} are uniformly distributed in the network area, i.e., $\vv{u}_k \sim \mathcal{U}(A)$. To show the robustness of \name{} in realistic propagation scenarios, we relax the assumption of \gls{los} propagation conditions and account for the \gls{nlos} paths as well.
\change{
In particular, we consider the stochastic geometry-based model in~\cite{gapeyenko2017temporal}, which models the blockers as cylinders of height $h_{\ssub{B}}$, and diameter $r_{\ssub{B}}$ distributed according to a \gls{ppp} with intensity $\lambda_{\ssub{B}}$.
The pathloss exponent for the \gls{los} or \gls{nlos} paths are denoted by $\chi_{\text{LoS}}$ or $\chi_{\text{NLoS}}$, respectively.
}
\change{We consider a frame duration $T$ of $10$ ms, divided in a total of $10$ time slots comprising one slot for channel estimation and probing phase, $N_D=8$ \gls{ul}, and $N_U=3$ \gls{dl} time slots.}
As far as the optimization of the precoder at the \gls{bs} is concerned, we assume perfect \gls{csi} at the \gls{bs}. In particular, the configuration of the \gls{hris} is assumed to be fixed by \name{} when optimizing the \gls{bs} precoder.
Therefore, the system is equivalent to a \gls{miso} channel given by the sum of the direct and reflected paths between the \gls{bs} and each \gls{ue}, where the \gls{hris} is viewed as an additional fixed scatterer (whose optimization is obtained by using \name{}). For a fair comparison with the benchmark scheme in \cite{Mursia2021}, the \gls{bs} precoder is chosen as
\begin{align}
    \vv{W}=\sqrt{P}\frac{\left( \vv{H}\vv{H}^{\herm} + \mu \vv{I}_m\right)^{-1}\vv{H}}{\|\left( \vv{H}\vv{H}^{\herm} + \mu \vv{I}_m\right)^{-1}\vv{H}\|_F},
\end{align}
where $\mu = K\frac{\sigma^2_n}{P}$, and each column of $\vv{H} \in \mathbb{C}^{M \times K}$ is the equivalent end-to-end \gls{miso} channel between the \gls{bs} and the corresponding \gls{ue}. 

It is worth mentioning that the performance of centralized deployments and \name{} depends on the overhead for channel estimation and reporting \cite{zappone2020overhead}, and the overhead of the probing phase \cite{rouissi2015design}, respectively. These two solutions are very different from each other and a fair comparison of the associated overhead is postponed to future research work.

\vspace{-0.3cm}
\subsection{Comparison with Centralized Deployment}

\begin{figure}[t]
        \center
        \includegraphics[width=0.9\linewidth, trim = {0cm, 4cm, 0cm, 0cm}]{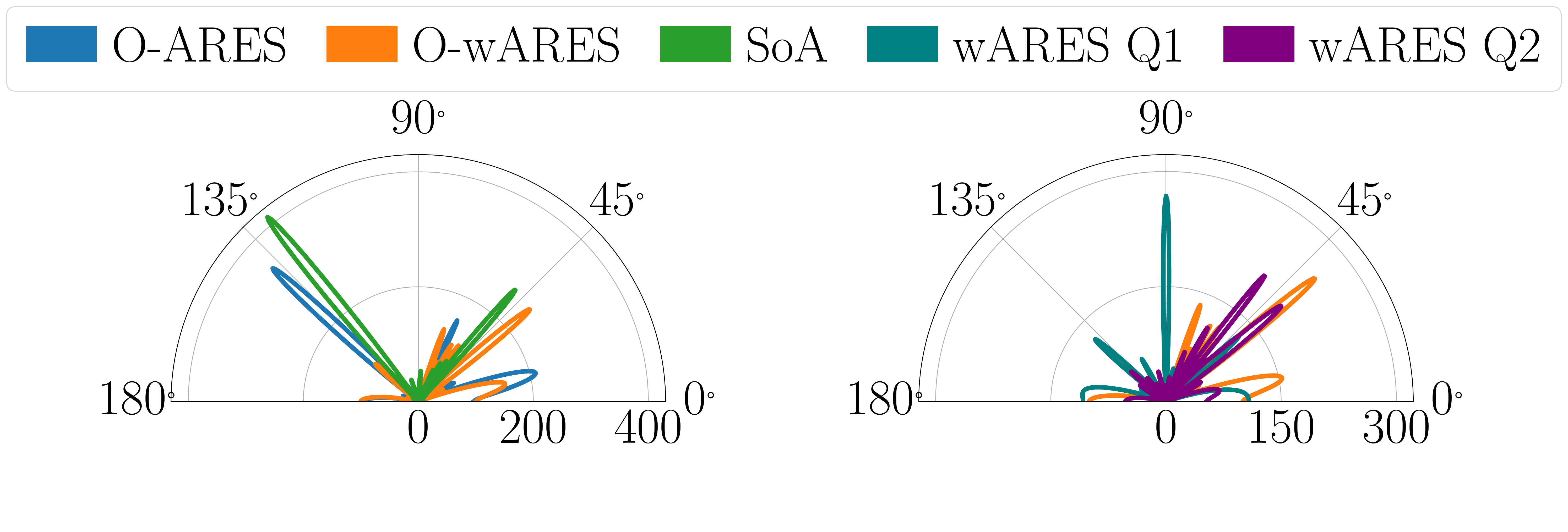}
        \vspace{-0.2cm}\caption{\label{fig:beam_pattern_marisa}
        Radiation pattern at the \gls{hris} along the azimuth directions obtained with O-\name{}, O-w\name{}, and \gls{soa}~\cite{Mursia2021} (left), and radiation patterns obtained with $Q$ bits of phase quantization (right).}
\end{figure}

\begin{figure}[t]
        \center
        \includegraphics[width=0.9\linewidth]{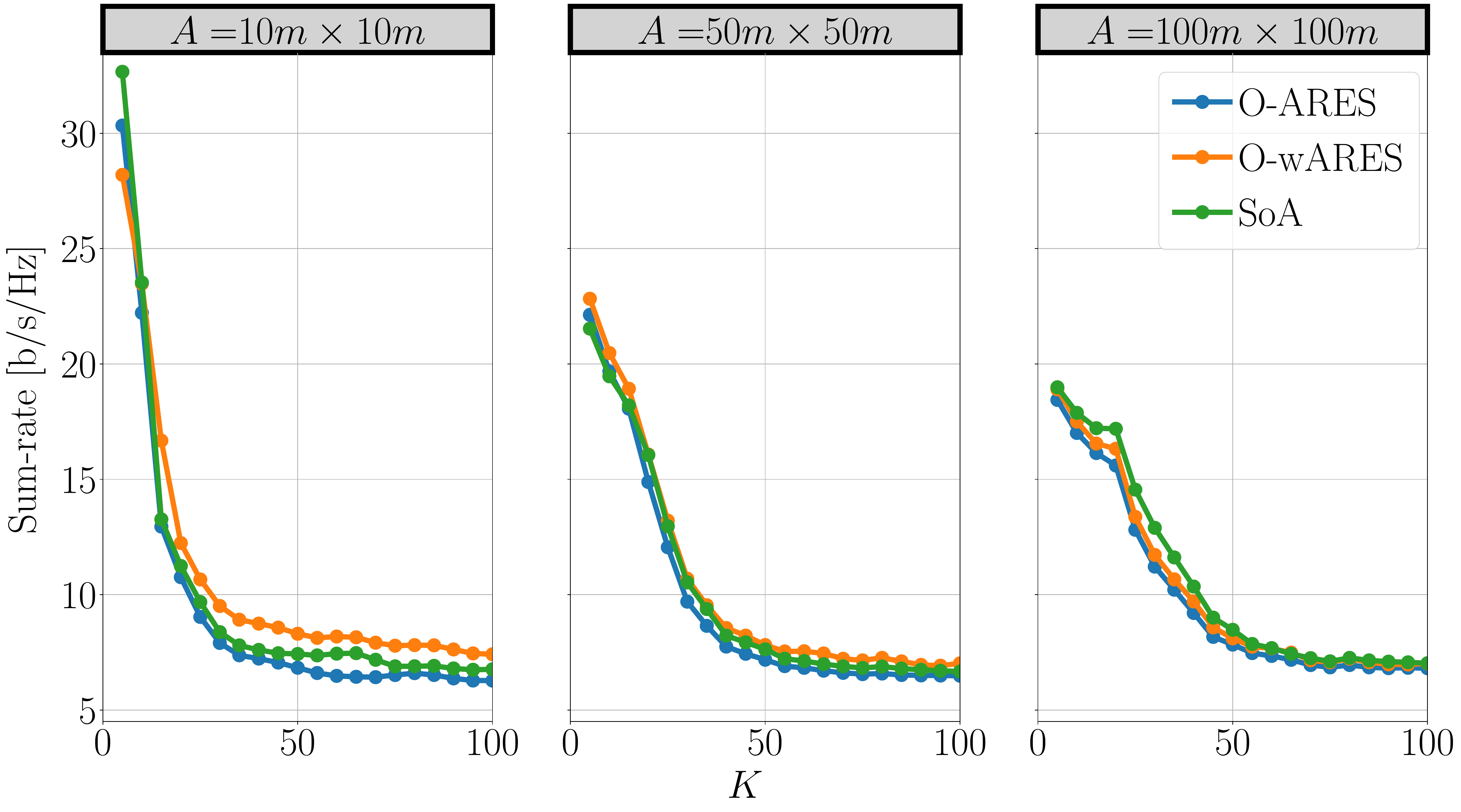}
        \vspace{-0.2cm}\caption{\label{fig:oracle_vs_SoA_K} Average sum-rate in a multi-\gls{ue} scenario obtained by solving Problem~\ref{problem:max_sinr_multi_ue} with perfect \gls{csi} and by \gls{soa}~\cite{Mursia2021} against the number of \glspl{ue} $K$ for different network areas and when the number of \gls{hris} elements is $N=32$.}
\end{figure}

We analyze the viability of self-configuring an \gls{hris} by solving Problem~\ref{problem:max_sinr_multi_ue} with perfect knowledge of the aggregate \gls{hris}-\gls{ue} channel $\vv{h}_{\ssub{\Sigma}}$ and of the response vector of the \gls{hris} towards the \gls{bs} $\vv{a}_{\ssub{R}}(\vv{b})$ in~\eqref{eq:v_BU}. We refer to this design as the \textit{Oracle} (O) scheme, since the channels are assumed to be perfectly known. 
Moreover, we analyze two solutions that assume real-valued (continuous) phase shifts: $i$) O-\name{}, which calculates $\vv{h}_{\ssub{\Sigma}} = \sum_k \frac{\vv{h}_{k}}{\|\vv{h}_{k}\|}$, and $ii$) O-weighted \name{} (O-w\name{}), which calculates $\vv{h}_{\ssub{\Sigma}} = \sum_k \vv{h}_{k}$. Specifically, O-\name{} estimates $\vv {v}_{\ssub{B}}$ and $\vv{\uptheta}_{\ssub{U}}$ only based on the directions of the paths that are assumed to have a unit gain, while O-w\name{} utilizes the direction and the gain of the paths.

Figure~\ref{fig:beam_pattern_marisa} (left) shows a comparison of the \gls{hris} configuration obtained by O-\name{}, O-w\name{}, and the \gls{soa} centralized solution in~\cite{Mursia2021}, which jointly optimizes the \gls{bs} precoder and the \gls{ris} phase shifts by means of a control channel.
While the \gls{soa} provides a very directive beampattern with few enhanced directions, both versions of O-\name{} result in a wider range of directions at the expense of a smaller gain due to the presence of multiple secondary lobes. Despite the different beampatterns, the sum-rates obtained by \name{} and the centralized benchmark, as shown in Fig.~\ref{fig:oracle_vs_SoA_K}, are very similar. In particular, O-\name{} and the \gls{soa} provide a sum-rate that does not increase with the number of \glspl{ue}, which hints at an interference-constrained scenario. Notably, O-w\name{} delivers better performance thanks to the weighting mechanism that strengthens the reflected paths with higher power gains. 

\vspace{-0.25cm}
\subsection{Codebook-Based \name{}}

In this section, we analyze the performance offered by \name{} under the realistic assumption that the \gls{hris} optimizes its configuration through power measurements and by iteratively activating the beam patterns (codewords) in the codebook $\mathcal{C}$. Therefore, no apriori knowledge of the aggregate channel $\vv{h}_{\ssub{\Sigma}}$ and of the response vector towards the \gls{bs} $\vv{a}_{\ssub{R}}(\vv{b)}$ is assumed. Also, the phase shifts applied by the \gls{hris} belong to the discrete set $\mathcal{Q}$ in \eqref{eq:phase}.
The steering directions are computed based on the estimated peaks in the measured power profile $\mathcal{P}$.
Supported by the previous case study, we analyze only the performance of w\name{}. Based on the estimated angular power profile  $\mathcal{P}$, $\vv{\upphi}_{\ssub{B}}$ and $\vv{\upphi}_{\ssub{U}}$ are estimated from \eqref{eq:v_B_code}. The weights $\delta_l$ are set equal to $\rho_l$,  $\forall l \in \mathcal{I}$ and $0$, $\forall l \in \hat{\mathcal{I}}$.

\begin{figure}[t]
        \center
        \includegraphics[width=.9\linewidth]{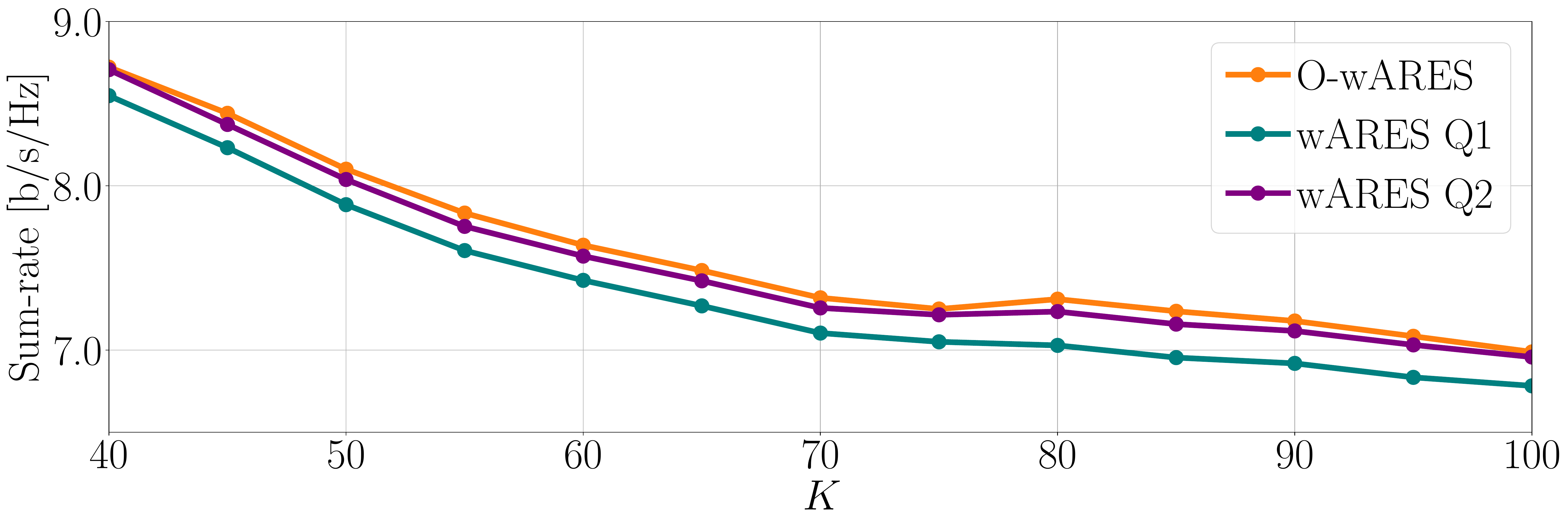}
        \vspace{-0.2cm}\caption{\label{fig:oracle_vs_est_K}
        Average sum-rate in a multi-\gls{ue} scenario obtained by w\name{} for different quantization levels $Q$, and O-w\name{}, against the number of \glspl{ue} $K$.}
\end{figure}
\begin{figure}[t]
        \center
        \includegraphics[width=.9\linewidth]{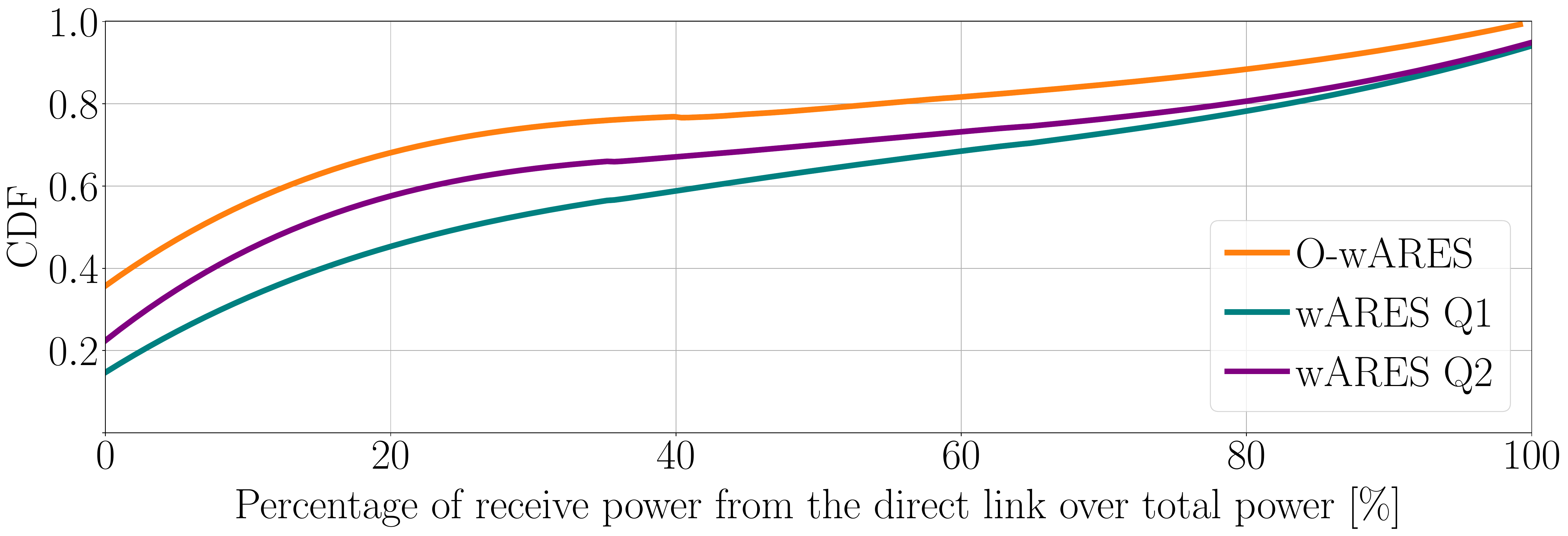}
        \vspace{-0.2cm}\caption{\label{fig:pdf} Cumulative distribution function (CDF) of the fraction of the receive power at each \gls{ue} over the direct path with respect to the total receive power after precoder optimization and selection at the \gls{bs}.}
\end{figure}

We consider two implementations for w\name{}, denoted by w\name{} Q1 and w\name{} Q2, which correspond to the w\name{} algorithm with the quantization levels $Q=1$ and $Q=2$ bits, respectively. The achievable average sum-rate is reported in Fig.~\ref{fig:oracle_vs_est_K}. Relaxing the assumptions of perfect \gls{csi} and continuous phase shifts has only a limited impact on the sum-rate, which confirms the effectiveness of the proposed approach proposed. As expected, the sum-rate worsens when one quantization bit is used, while two quantization bits offer good performance already.

In Fig.~\ref{fig:pdf}, finally, we report the distribution of the percentage of power that every \gls{ue} receives from the direct link with respect to the total received power (from the direct link and the reflected link). We note that O-w\name{} offers the highest power boost that originates from the reflect paths thanks to its ideal beamforming capabilities. On the other hand,  w\name{} Q$1$ and Q$2$ are affected by quantization errors that lead to  beampatterns with a more distributed power spread. Similar unwanted reflections can be seen in Fig.~\ref{fig:beam_pattern_marisa} (right), where the beampatterns obtained with O-w\name{} and w\name{} Q$1$ and Q$2$ are reported. However, these unwanted reflections minimally affect the sum-rate.

\vspace{-0.3cm}
\subsection{Energy Self-sufficiency}

\begin{figure}[t]
        \center
        \includegraphics[width=.9\linewidth]{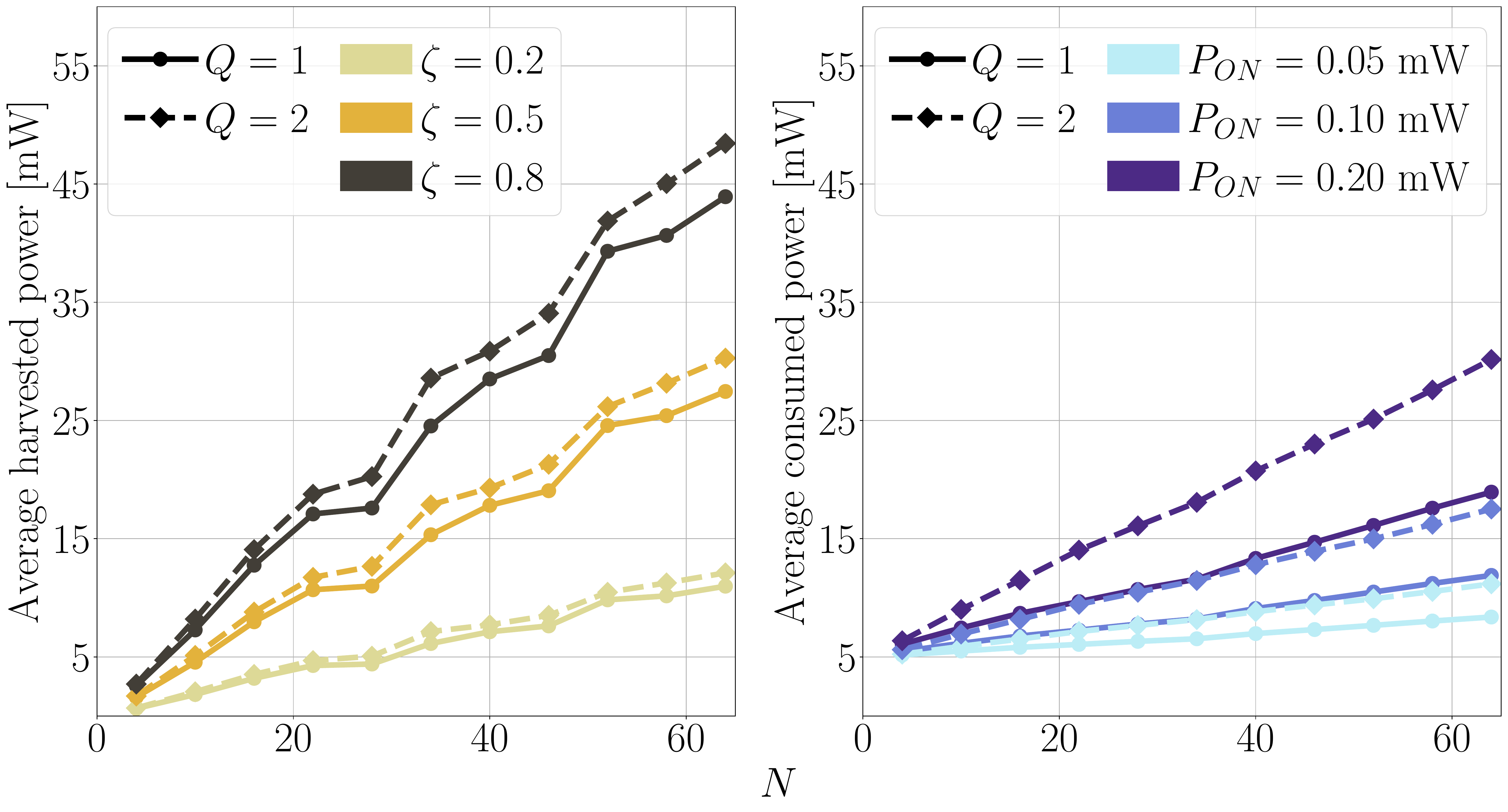}
        \vspace{-0.5cm}\caption{\label{fig:harvesting_performances} {\name{} average harvested and consumed power against the number of \gls{hris} elements $N$ for different values of traffic $\zeta$, \gls{pin} diode activation consumption $P_{ON}$, and phase quantization $Q$. The network area is $A = 50m \times 50m$, and the number of user is $K = 75$.}}\vspace{-0.2cm}
\end{figure}

\begin{figure}[t]
        \center
        \includegraphics[width=.9\linewidth]{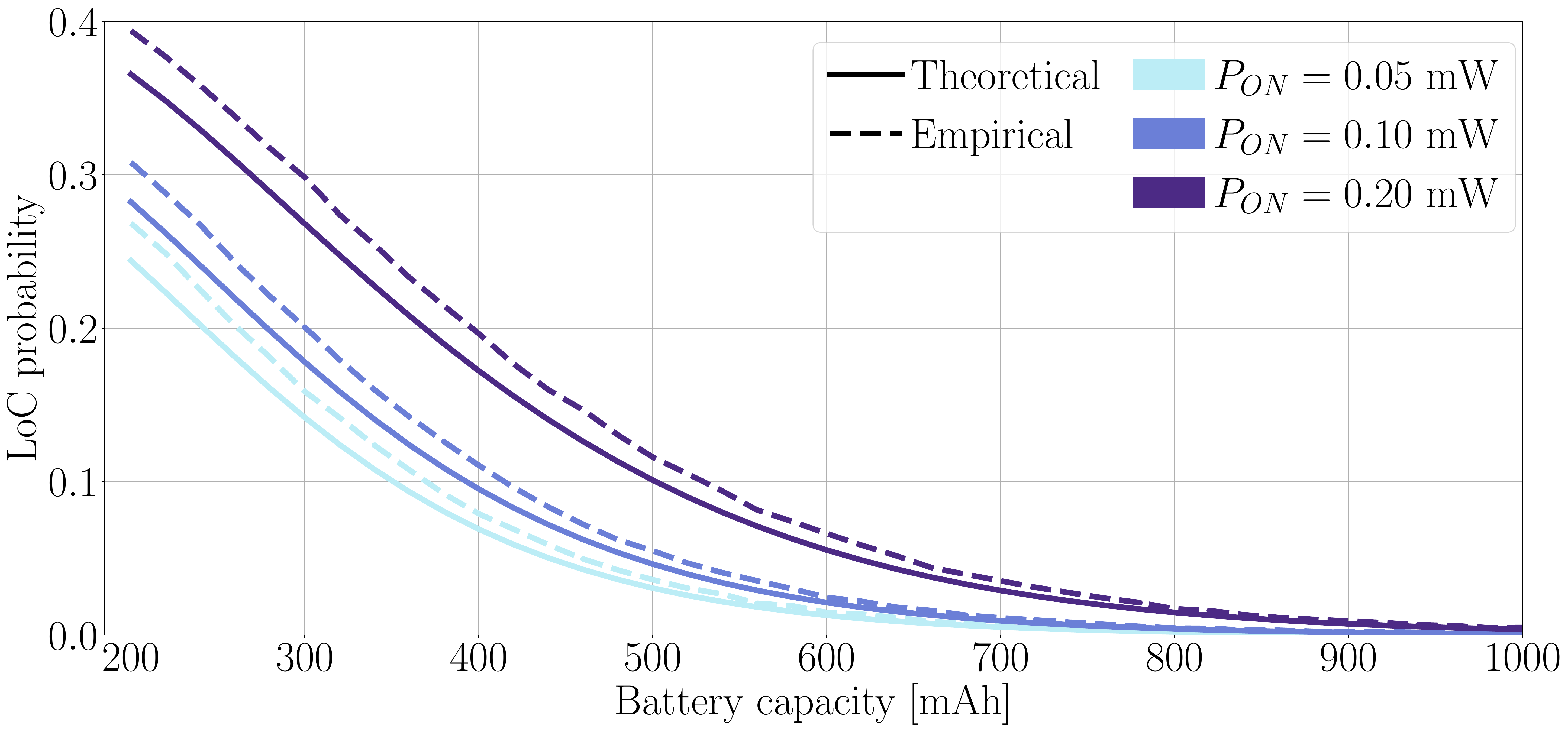}
        \vspace{-0.4cm}\caption{\label{fig:battery_sizing}{\name{} probability against the battery capacity. The \gls{hris} has $N = 40$ meta-atoms with quantization $Q=2$ and the scenario includes $K = 75$ users with traffic $\zeta = 0.5$.}}\vspace{-0.1cm}
\end{figure}

\begin{figure}[t]
        \center
        \includegraphics[width=1\linewidth]{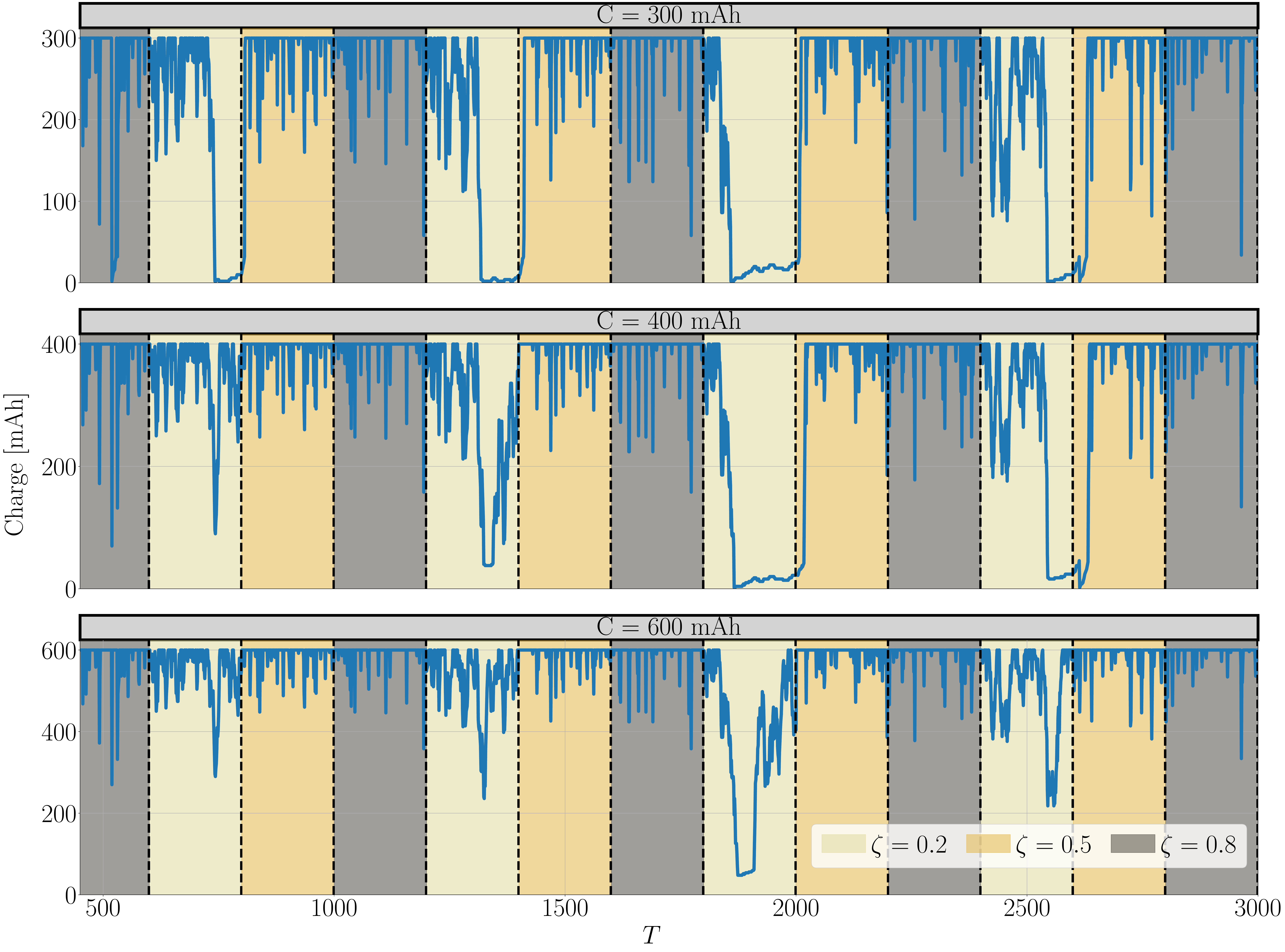}
        \vspace{-0.8cm}\caption{\label{fig:battery_charge_evolution}{Example of charge/discharge process obtained with different capacity of the battery, considering an \gls{hris} with $N = 40$ meta-atoms, quantization $Q=2$, and $P_{ON} = 0.1$ mW, deployed in a with $K = 75$ users, and $\zeta = \{0.2, 0.5, 0.8\}$ in yellow, cyan, and red respectively.}}
\end{figure}

\change{
Lastly, we assess the performances of \name{} in terms of energy self-sufficiency, starting from the evaluation of the harvested and consumed power, and considering a low \gls{soc} threshold $\Gamma=10\%$. Fig.~\ref{fig:harvesting_performances} shows the average harvested (left) and consumed (right) power with different \gls{hris} hardware configurations, and in different traffic conditions. From these results, we derive the statistics of $\Delta E$, assuming that it is distributed as a Gaussian \gls{rv}, i.e., $\Delta E \sim \mathcal{N}(\mu_{\Delta},\,\sigma^{2}_{\Delta})$. Moreover, to take into account the power consumption of the \gls{hris} controller, we consider the absorption of $4.9$ mW and $1.8$ mW when the device operates in standard or idle condition, respectively\footnote{\change{The selected values correspond to the power consumption of the run and the stop modes of the STM32L071V8T6 micro controller adopted in the \gls{ris} implementation proposed in~\cite{rossanese2022}.}}.
The average harvested power has a monotonic increasing behavior concerning the number of meta-atoms composing the surfaces, as the larger the number of elements, the higher the \gls{hris} ability to focus towards the signal sources, which, in turn, grows the harvested power.
Similarly, more quantization levels $Q$ allow for finer granularity in the phase shift selection, further improving the harvesting performance. However, higher values of $N$ and $Q$ increase the complexity of the \gls{hris} hardware, inflating the corresponding overall power consumption. As a result, a novel trade-off arises between the harvested and the consumed powers, whose deviation needs to be compensated by the battery in order to keep the \gls{hris} alive with the desired $p_{LoC}$.

In Fig.~\ref{fig:battery_sizing}, we depict the $p_{LoC}$ for different sizes of the device battery, and different values of $P_{ON}$, which directly affects the \gls{hris} power consumption. We consider both the $p_{LoC}$ obtained as per \eqref{eq:p_loc}, denoted as theoretical, and by simulating the charge and discharge processes over $10^7$ reconfiguration periods $T$, denoted as empirical. Besides, we consider a battery voltage of $3.7$ V and set $\Delta$ to $10$ mAh. As expected, increasing the battery size reduces the \gls{loc} probability. However, for high values of $P_{ON}$, i.e. high overall power consumption, enlarging the size of the battery might not provide adequately low $p_{LoC}$ as the harvested power becomes smaller than the overall \gls{hris} hardware consumption. Fig.~\ref{fig:battery_charge_evolution} showcases an example of the battery charge and discharge process for different traffic intensities $\zeta \in  \{0.2, 0.5, 0.8\}$. In particular, we illustrate how the battery tends to discharge in low-traffic conditions and recharge in higher-traffic conditions. We also confirm that equipping the \gls{hris} with a bigger battery is beneficial for avoiding the \gls{loc} as it helps the \gls{hris} not run out of battery during prolonged low-traffic conditions.}

\vspace{-0.3cm}
\section{Conclusions}
\label{sec:conclusion}
\change{\glspl{ris} are an emerging clean-slate technology with the inherent potential of fundamentally reshaping the design and deployment of mobile communication systems. In this paper, we introduced \name{}, an Autonomous RIS with Energy
harvesting and Self-configuration solution towards 6G, which dismisses the need for a complex ad-hoc control channel and a power supply to operate \glspl{ris}, bringing up a \emph{fully-autonomous} \glspl{ris} solution with no deployment constraint. \name{} is built upon $i$) a new low-complexity hardware design providing \glspl{hris} with both sensing and energy harvesting capabilities, $ii$) a channel estimation model lato-sensu at the \glspl{hris}, $iii$) an autonomous \glspl{hris} configuration methodology operating both on the reflection and on the \gls{eh} properties of the \glspl{hris}, which is based only on locally estimated \gls{csi}, without a control channel.
\name{} achieves communication perfomance comparable with fully \gls{csi}-aware benchmark while demonstrating the feasibility of \gls{eh} for \glspl{ris} power supply in future deployments.}

\vspace{-0.3cm}
\bibliographystyle{IEEEtran}
\bibliography{IEEEabrv, references_short}

\end{document}